\documentstyle[emulateapj,epsf,apjfonts]{article}

\slugcomment{Submitted to ApJ}

\begin{document}

\title{Evolution of the Solar Activity over Time and Effects on
Planetary Atmospheres: I. High-energy Irradiances (1--1700 \AA)}

\author{Ignasi Ribas\altaffilmark{1}, Edward F. Guinan\altaffilmark{2},
Manuel G\"udel\altaffilmark{3}, and Marc Audard\altaffilmark{4}}

\altaffiltext{1}{Institut d'Estudis Espacials de Catalunya/CSIC, Campus UAB,
Facultat de Ci\`encies, Torre C5 - parell - 2a planta, 08193 Bellaterra, Spain;
E-mail: iribas@ieec.uab.es}

\altaffiltext{2}{Department of Astronomy \& Astrophysics, Villanova University,
Villanova, PA 19085, USA; E-mail: edward.guinan@villanova.edu}

\altaffiltext{3}{Paul Scherrer Institut, W\"urenlingen \& Villigen, 5232
Villigen PSI, Switzerland; E-mail: guedel@astro.phys.ethz.ch}

\altaffiltext{4}{Columbia Astrophysics Laboratory, Columbia University, 550
West 120th Street, New York, NY 10027, USA; E-mail: audard@astro.columbia.edu}

\begin{abstract}
We report on the results of the Sun in Time multi-wavelength program
(X-rays to the UV) of solar analogs with ages covering $\sim$0.1--7 Gyr.
The chief science goals are to study the solar magnetic dynamo and to
determine the radiative and magnetic properties of the Sun during its
evolution across the main sequence. The present paper focuses on the
latter goal, which has the ultimate purpose of providing the spectral
irradiance evolution of solar-type stars to be used in the study and
modeling of planetary atmospheres. The results from the Sun in Time
program suggest that the coronal X-ray--EUV emissions of the young
main-sequence Sun were $\sim$100--1000 times stronger than those of the
present Sun. Similarly, the transition region and chromospheric FUV--UV
emissions of the young Sun are expected to be 20--60 and 10--20 times
stronger, respectively, than at present. When considering the integrated
high-energy emission from 1 to 1200~\AA\, the resulting relationship
indicates that the solar high-energy flux was about 2.5 times the present
value 2.5 Gyr ago and about 6 times the present value about 3.5 Gyr ago
(when life supposedly arose on Earth). The strong radiation emissions
inferred should have had major influences on the thermal structure,
photochemistry, and photoionization of planetary atmospheres and also
played an important role in the development of primitive life in the Solar
System. Some examples of the application of the Sun in Time results on
exoplanets and on early Solar System planets are discussed.
\end{abstract}


\section{Introduction}

The Sun is by far the most important star to us. Without a dependable
(stable) star like the Sun, the Earth would not have developed a rich and
diverse biosphere that is home to millions of species of life. Because of
ever accelerating nuclear reactions in its core, the Sun is a slowly
evolving variable star that has undergone a $\sim$40\% increase in
luminosity over the last 4.5 Gyr, as predicted by the standard solar
evolution model (e.g., Girardi et al. 2000). On much shorter timescales,
we know that the Sun is also a magnetic variable star, with an $\sim$11-yr
sunspot and activity cycle and a $\sim$22-yr magnetic cycle. As predicted
by magnetic dynamo theory, the Sun's rotation ($P_{\rm rot} \sim 25.5$ d)
and convective outer envelope interact to generate magnetic fields (Parker
1970). The magnetic dynamo-generated energy is released in the form of
flares and chromospheric, transition-region, and coronal radiation.
Satellite observations of the Sun show that it undergoes small
(0.10--0.15\%) variations in total brightness over its activity cycle,
although the changes at short wavelengths (UV to X-rays) are much more
pronounced (10--500\%) (Lean 1997).

The magnetic activity of the present Sun is feeble relative to other
solar-type stars, but magnetic-induced phenomena still have important
effects on Earth and the Solar System (see Guinan \& Ribas 2002). Thus,
the fundamental question whether the Sun has always been a relatively
inactive star or if, in contrast, it has experienced some periods of
stronger magnetic activity has a strong impact on the evolution of the
Solar System. Compelling observational evidence (G\"udel et al. 1997a)
shows that zero-age main sequence (ZAMS) solar-type stars rotate over 10
times faster than today's Sun. As a consequence of this, young solar-type
stars, including the young Sun, have vigorous magnetic dynamos and
correspondingly strong high-energy emissions. From the study of solar type
stars with different ages, Skumanich (1972), Simon et al. (1985), and
others showed that the Sun loses angular momentum with time via magnetized
winds (magnetic braking) thus leading to a secular increase of its
rotation period (Durney 1972). This rotation slow-down is well fitted by a
power law roughly proportional to $t^{-1/2}$ (e.g., Skumanich 1972;
Soderblom 1981; Ayres 1997). Note that the age--rotation period
relationship is tighter for intermediate/old stars, while young stars (a
few $10^8$ yr) show a larger spread in rotation periods. In response to
slower rotation, the solar dynamo strength diminishes with time causing
the Sun's high-energy emissions also to undergo significant decreases.
Comprehensive studies on this subject were published by Zahnle \& Walker
(1982) and Ayres (1997). The reader is referred to these publications for
background information on solar-type stars' upper atmospheres and related
high-energy emissions.

The ``Sun in Time'' program was established some 20 years ago (Dorren \&
Guinan 1994; Guinan \& Ribas 2004) to study a sample of
accurately-selected solar proxies (G0--G5 V stars) with different ages
across the electromagnetic spectrum. The primary aims of the program are:
{\em 1)} To test dynamo models of the Sun in which rotation is the only
significant variable parameter, and {\em 2)} to determine the spectral
irradiance of the Sun over its main-sequence lifetime. In the present
paper, we focus on the latter goal, which has the ultimate purpose of
characterizing the evolution of the solar emissions with direct
application to the study and modeling of atmospheres of both Solar System
planets and exoplanets in orbit around solar-type stars. There are a
number of aspects within the Solar System in which a stronger high-energy
flux from the young Sun could have had a critical impact. Through
photochemical and photoionization processes, the strong X-ray and UV
emissions of the young Sun could have had a major effect on the evolution
of the atmospheres, ionospheres and climates of the terrestrial planets,
including the Earth (e.g., Canuto et al. 1982, 1983; Kasting \& Catling
2003; Smith et al. 2004). For example, paleo-climate models of the Earth
should account for the higher levels of ionizing and dissociating UV
radiation in the past.  Even the development of life on Earth (and
possibly on Mars) could have been influenced by the larger doses of
sterilizing UV radiation expected from the young Sun (cf. Cockell et al.
2000).

In this paper we present the results of an investigation on the long-term
magnetic history of the Sun, focused on the high-energy emissions (below
1700~\AA). Basically, we study the chromospheric, transition region and
coronal emissions, associated to high-temperature atmospheric layers. In
contrast, the time evolution of the photospheric emissions is already well
characterized because these scale with the overall bolometric luminosity.
We have made use of our selected sample of stellar proxies with ages that
cover most of the main-sequence lifetime of the Sun. A large number of
multiwavelength observations (X-ray, EUV, FUV, UV; hereafter XUV) of the
solar analogs have been collected to fully describe their spectral
irradiances as a function of age and rotation. Also discussed are the
major effects that the young Sun's strong XUV radiation may have had on
the photoionization, photochemistry, and erosion of paleo-planetary
atmospheres.

\section{The Sun in Time sample} \label{sec:targ}

A critical element to any study of the evolution of the Sun's irradiance
with time is a carefully-selected sample of stars to serve as proxies for
the Sun with different rotation periods, and therefore different ages. The
Sun in Time sample contains single or widely separated binary, nearby,
G0--5 stars that have known rotation periods and well-determined
temperatures, luminosities, and metallicities. In addition, we have been
able to estimate the stellar ages by making use of their memberships in
clusters and moving groups, rotation period--age relationships, and, for
the older stars, fits to stellar evolution models. Comparisons with
stellar models predict stellar masses within 10\% of 1~M$_{\sun}$. While
the complete Sun in Time sample contains over 15 solar analogs, we focus
here on six stars that have been observed with a variety of high-energy
instruments: EK Dra, $\pi^1$ UMa, $\chi^1$ Ori, $\kappa^1$ Cet, $\beta$
Com, and $\beta$ Hyi. These solar analogs cover most of the Sun's main
sequence lifetime at approximate ages of 100 Myr, 300 Myr, 650 Myr, 1.6
Gyr, and 6.7 Gyr. Note that we have not used any proxy for the current Sun
but the Sun itself. Although there are no full-disk high-resolution
spectra of the Sun (as pointed out by Pagano et al. 2004), the datasets
described in \S \ref{sec:solar} have sufficient resolution to fulfill the
requirements of the present study. If higher resolution data were
necessary, a valid alternative would be to use solar twins. For example,
18 Sco is a nearly perfect solar twin (Porto de Mello \& da Silva 1997;
Hamilton et al. 2003), but few high-energy observations are available
because of its relative faintness. $\alpha$ Cen A is slightly more massive
than the Sun but yet a good solar twin (Pagano et al. 2004) and with
numerous observations. In this case, however, some of the high-energy
observations include the emissions from the active K-type companion
$\alpha$ Cen B, which complicates the analysis significantly.

A discussion of each observed target is provided below and a summary of
the relevant stellar data (including those for the Sun) is shown in Table
\ref{tabtarg}. Stellar radii have been estimated from the observed
magnitude, distance, and temperature, while masses generally follow from
evolutionary model calculations. Also given in Table \ref{tabtarg} is the
estimated value of the solar radius at the corresponding age as provided
by the stellar models of Girardi et al. (2000).

{\bf EK Dra:} This is a nearby G1.5~V star that has traditionally been
considered among the most active solar analogs in our neighborhood. Its
main properties were reviewed by Strassmeier \& Rice (1998) and Messina \&
Guinan (2003), including an average rotation period of 2.68 days.
Evolutionary models of Girardi et al. (2000) yield a slightly super-solar
mass and a ZAMS age at the observed temperature, luminosity and chemical
composition (solar) of EK Dra. Montes et al. (2001a,b), and references
therein, classified EK Dra as a kinematic member of the so-called Local
Association or Pleiades moving group (with an estimated age range of
20--150 Myr) but inferred an age younger than the Pleiades cluster from
the observed Li abundance. Similar conclusions were drawn by Wichmann et
al. (2003), who obtained an upper limit to the age of EK Dra of 50--100
Myr. Here we adopt an age of $\sim$100 Myr, which seems to be a good
compromise. Note that EK Dra was found to be a radial velocity variable by
Duquennoy \& Mayor (1991) with a period of about 12.5 yr. This companion
to EK Dra would have a minimum mass below 0.4~M$_{\sun}$ and and thus
likely be an M-type star. Metchev \& Hillenbrand (2004) have recently
reported the discovery of a companion to EK Dra with a mass of
0.20$^{+0.30}_{-0.08}$~M$_{\sun}$ using adaptive optics. It is not yet
clear whether these two low-mass companions detected independently are
indeed the same. In any case, neither of them is a concern for our studies
since they should only contribute a few percent in the wavelength domain
of interest.

{\bf $\pi^1$ UMa:} This young solar proxy is an active G1.5~V star with a
rotation period of about 4.9 days (Messina \& Guinan 2003). Gaidos \&
Gonz\'alez (2002) and Ottmann et al. (1998) carried out detailed
spectroscopic analyses and determined accurate values for the stellar
temperature and metal abundance (compatible with solar). Evolutionary
models of Girardi et al. (2000) indicate a mass just a few percent higher
than that of the Sun. Kinematic studies such as those of Montes et al.
(2001a,b) and King et al. (2003) classify $\pi^1$ UMa as a probable member
of the Ursa Major moving group.  Although the age of this kinematic group
of stars has traditionally been quoted as about 300 Myr (see, e.g.,
Soderblom \& Mayor 1993), recent work by King et al. (2003) suggests an
older age of 500$\pm$100 Myr on the basis of a number of complementary
criteria. Here we prefer to adopt the canonical age of $\sim$300 Myr since
it stands in much better agreement with our measured high-energy fluxes as
discussed below.

{\bf $\chi^1$ Ori:} This field G1~V star has a rotation period of about
5.2 days (Messina et al. 2001), which is indicative of its young age. Its
temperature and metal abundance (slightly sub-solar) were determined by
Gratton et al. (1998) and Taylor (2003a,b) with very similar results. King
et al. (2003) classified $\chi^1$ Ori as a certain member of the Ursa
Major moving group in agreement with a number of previous studies. In
accordance with $\pi^1$ UMa, we have adopted an age estimate of $\sim$300
Myr. Interestingly, $\chi^1$ Ori is a spectroscopic binary with an orbital
period of about 14 yr (Han \& Gatewood 2002) that was recently resolved
using adaptive optics at Keck by K\"onig et al. (2002). These authors were
able to determine the dynamical mass of both $\chi^1$ Ori and its
companion. Note that, with a mass of only 0.15~M$_{\sun}$, the quiescent
high-energy emissions of $\chi^1$ Ori~B will be negligibly small compared
to those of its much larger primary companion.

{\bf $\kappa^1$ Cet:} With spectral type G5~V, this star the coolest in
the sample. Spectroscopic parameters were determined by Gaidos \&
Gonz\'alez (2002), who also estimated slightly super-solar metal content.
Its rotation period was reported to be around 9.2 days by Messina \&
Guinan (2003), in agreement with other determinations. Evolutionary models
of Girardi et al. (2000) yield a mass very close to the solar value. In
the kinematic study of Montes et al. (2001a) $\kappa^1$ Cet was not
flagged as a member of any of the canonical stellar groups but was
classified as a young disk star. In the absence of more direct indicators,
we estimate the age of $\kappa^1$ Cet from its rotation period and mean
X-ray luminosity ($\log L_{\rm X}=28.8$ [erg s$^{-1}$]; G\"udel et al.
1997a). Comparison with Hyades stars in the same (B-V) interval as
$\kappa^1$ Cet reveals that both its rotation period (cf. Radick et al.
1995) and its X-ray luminosity (cf. Barrado y Navascu\'es et al.  1998)
are close to the average of the Hyades members. These criteria suggest for
$\kappa^1$ Cet an age close to the canonical Hyades age of $\sim$650 Myr,
which we subsequently adopt. Note that our value is older than the age
estimated by Lachaume et al. (1999).

{\bf $\beta$ Com:} This is a G0~V star with a rotation period of about 12
days (Gray \& Baliunas 1997). Temperature and metal abundance (slightly
super-solar) were determined by Barklem et al. (2002) and Gray et al.
(2001) from spectroscopic analyses. Using the observed data, the
evolutionary models of Girardi et al. (2000) yield a mass about 10\%
larger than the Sun and an age of 2.3$\pm$1.1 Gyr, in agreement with the
remark of Gray \& Baliunas (1997) about $\beta$ Com being younger than the
Sun. To refine the age estimate, we have made use of the rotation
period--age relationship for solar-type stars of Guinan et al. (1998) and
derived a value of $\sim$1.6 Gyr.

{\bf $\beta$ Hyi:} As pointed out by Dravins et al. (1998), $\beta$ Hyi,
of spectral type G2~IV, is our nearest subgiant star. A detailed study of
$\beta$ Hyi was published by Fernandes \& Monteiro (2003), who also review
determinations of stellar parameters, including the mass and age of
$\sim$6.7 Gyr. We adopt the values in that recent work.  The rotation
period of $\beta$ Hyi was determined from the analysis of 18 IUE High
Dispersion LWP spectra obtained during 1994/95 by Guinan under program
SPREG. Measures of the chromospheric Mg~{\sc ii} h\&k emission line fluxes
relative to the adjacent continuum were made. The analysis reveals an
apparent modulation in the relative Mg~{\sc ii} h\&k emission strength of
$P_{\rm rot}=29\pm3$ days arising from chromospheric faculae and plages
on the star (Guinan et al. 2005, in prep.).

There is an additional important issue yet to be addressed and this is the
interstellar medium (ISM) column density along the targets' lines of
sight. Although the stars in the sample are nearby and have negligibly
small values of $E(B-V)$ from ISM dust, some of the studied stellar
emission features suffer strong ISM absorption and appropriate corrections
need to be applied. The H~{\sc i} column densities in the lines of sight
of our targets were estimated from the various lines of sight sampled by
the H~{\sc i} Lyman $\alpha$ observations of Wood et al. (2004). For
$\kappa^1$ Cet and $\chi^1$ Ori direct measurements are available, whereas
for the other four stars we employed the measured H~{\sc i} column
densities for neighboring stars (selected on the basis of a similar
distance and position). We used HD 116956 for EK~Dra, DK~UMa for
$\pi^1$~UMa, HZ 43 for $\beta$~Com, and $\zeta$~Dor, $\epsilon$ Ind, and
HD 203244 for $\beta$~Hyi. The total adopted H~{\sc i} column densities in
the target lines of sight are given in Table \ref{tabtarg}. These can be
eventually scaled to compute the column densities of other elements from
mean local ISM abundances.

\section{Observational data} \label{sec:data}

Observations of the target stars in Table \ref{tabtarg} were carried out
with a variety of space-based instruments to maximize spectral coverage.
Data from the following space missions have been used in the present
study: {\em Advanced Satellite for Cosmology and Astrophysics} (ASCA),
{\em R\"ontgen Satellite} (ROSAT), {\em Extreme Ultraviolet Explorer}
(EUVE), {\em Far Ultraviolet Spectroscopic Explorer} (FUSE), {\em Hubble
Space Telescope} (HST), and {\em International Ultraviolet Explorer}
(IUE).  As can be seen in the summary presented in Table \ref{tabobs}, the
observations discussed in this paper cover approximately the interval
between 1~\AA\ and 1700~\AA, except for a gap between 360~\AA\ and
920~\AA, which is a region of very strong ISM absorption (H~{\sc i} Lyman
continuum), thus far largely unexplored for stars other than the Sun. The
number of datasets used for this study is quite extensive and thus all
observation identification files for each target and mission are listed in
Table \ref{tabobsid}. All the observations described here, both from our
guest observer programs and public datasets, were downloaded from the
HEASARC\footnote{http://heasarc.gsfc.nasa.gov/} and
MAST\footnote{http://archive.stsci.edu/} archives.

\subsection{X-rays: ASCA \& ROSAT}

\begin{figure}[!ht]
\epsscale{0.5}
\plotone{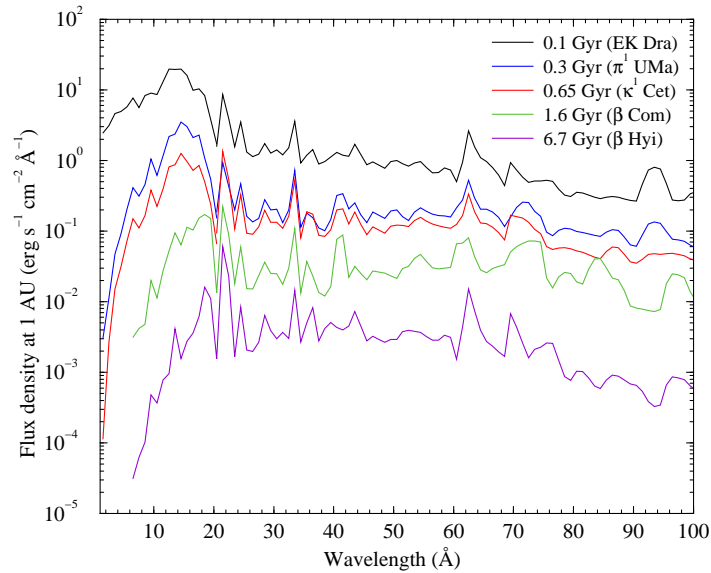}
\figcaption{X-ray spectral irradiances (flux density at 1 AU vs. wavelength)
covering different stages of the evolution of solar-type stars. The plot 
represents the fluxes in 1~\AA\ bins as predicted by multi-T plasma fits to 
ASCA and ROSAT observations (see text). Note the very large differences 
between young and old solar-type stars of up to a factor 1000.
\label{figxray}}
\end{figure}

A key aspect of the study is the transformation of the instrumental fluxes
into absolute fluxes, which is especially critical for the X-ray data. The
observations obtained with ASCA and ROSAT are not naturally in absolute
flux units and have to be compared with a physical plasma emission model
to perform the calibration. The ASCA observations were obtained with the
SIS0 and SIS1 detectors (Tanaka et al. 1994) and pointed observations with
the PSPC instrument (Briel \& Pfeffermann 1986) were used with ROSAT. We
reduced the X-ray data in the classical manner with the {\sc xselect}~V2.2
and {\sc xspec}~11.3.0 packages within {\sc ftools}. Then, as is commonly
done for coronal emissions (such as those from our solar analogs), we
considered a multi-$T_e$ plasma and ran a $\chi^2$ optimization fit with
the {\sc mekal} model (Mewe et al. 1995). In this procedure we followed
the exact same prescriptions as in G\"udel et al.  (1997a), and adopted
plasma models with 2 and 3 components (obtaining results entirely
consistent with this earlier study).  Then, absolute fluxes were
calculated from the best-fitting model with the aid of the {\sc xspec}
package. A correction for the H~{\sc i} column density -- using the values
given in Table \ref{tabtarg} -- was also included.  A plot comparing the
X-ray fluxes (ASCA and ROSAT) for the target stars is shown in Figure
\ref{figxray}, where large differences (of up to a thousand-fold) become
evident. Note that $\chi^1$ Ori only had ROSAT observations taken with 
the Boron filter (see footnote to Table \ref{tabobsid}. We decided not 
to use these here because of some calibration issues. G\"udel et al. 
(1997a), however, did analyze these data and obtained nearly the same 
parameters as for $\pi^1$ UMa (see their tables 3 and 5), as expected 
from their similar ages.

\subsection{EUV: EUVE}

\begin{figure}[!ht]
\epsscale{0.5}
\plotone{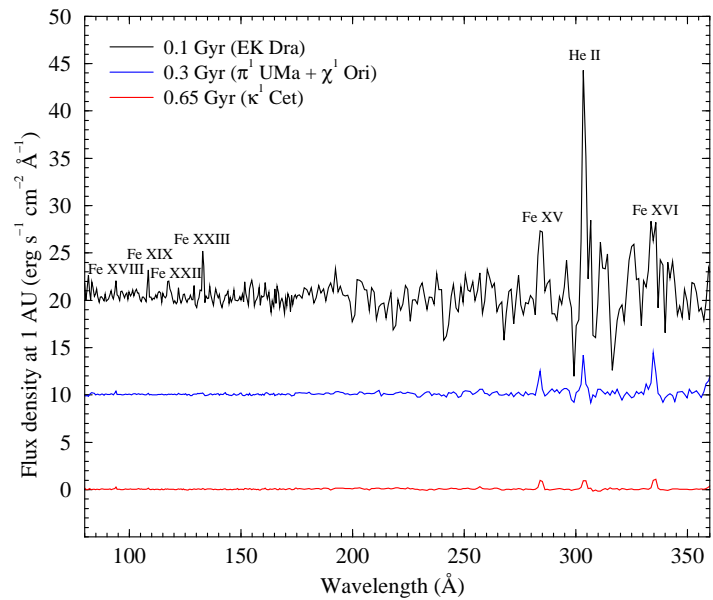}
\figcaption{Extreme UV irradiances (flux density at 1 AU vs. wavelength) 
covering different stages of the evolution of solar-type stars. Some 
relevant features are identified. The spectra have been zeropoint-shifted 
using integer multiples of 10 erg s$^{-1}$ cm$^{-2}$ \AA$^{-1}$ to avoid 
confusion. Note the decrease in emission line strength from top to bottom 
(i.e., increasing age and rotation period). 
\label{figeuve}}
\end{figure}

EUV observations of some of the targets were carried out as part of the
EUVE mission (Malina \& Bowyer 1991). In this case, the resulting data can
be directly flux calibrated during reduction and thus no emission model
has to be assumed beforehand. Data reduction of the EUVE spectra was
carried out following the same procedure as in G\"udel et al. (1997b).
The only detail worth mentioning here is the correction of the H~{\sc i}
column density, for which we assumed the values listed in Table
\ref{tabtarg}. An illustration of the EUV fluxes for the observed targets
is presented in Figure \ref{figeuve}. The spectra of $\pi^1$ UMa and
$\chi^1$ Ori, which correspond to the same stellar age, were averaged
together. Note that most of the output stellar flux is associated with
emission lines of highly-ionized element transitions.

\subsection{FUV: FUSE}

\begin{figure}[!ht]
\epsscale{0.5}
\plotone{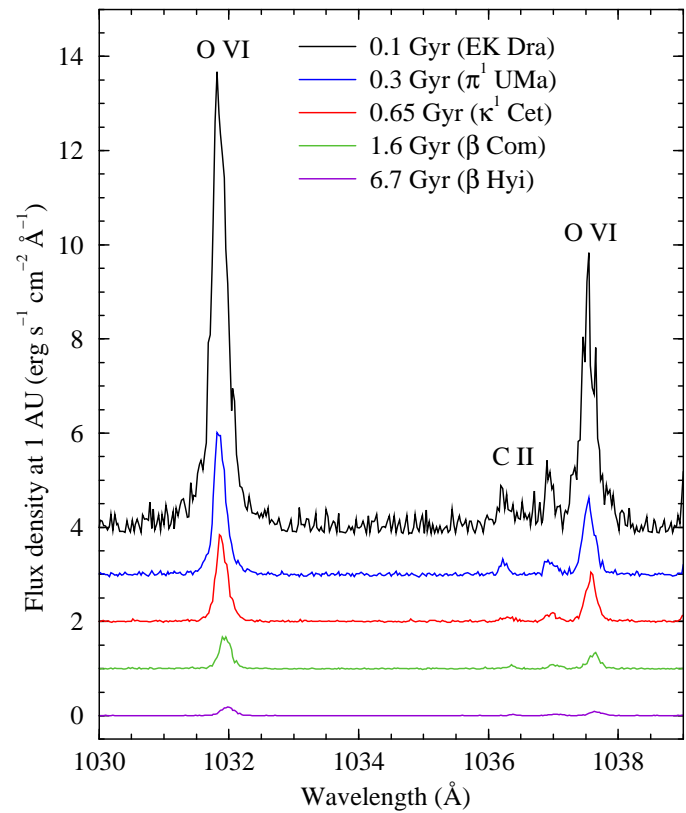}
\figcaption{Details of the far-UV irradiances for different stages of the
evolution of solar-type stars in the the region around the O~{\sc vi} 
$\lambda\lambda$1032,1038 doublet. The spectra have been zeropoint-shifted 
using integer multiples of 1 erg s$^{-1}$ cm$^{-2}$ \AA$^{-1}$ to avoid 
confusion. Note the obvious trend of decreasing flux with increasing stellar 
age. 
\label{figfuse}}
\end{figure}

To obtain irradiances in the FUV we carried out observations with FUSE
using its large aperture (Moos et al. 2000). For a detailed description of
the datasets used and the reduction procedure the reader is referred to
Guinan et al. (2003), and thus we shall skip the discussion here. The
reductions in Guinan et al. also included the correction of ISM absorption
in the emission features whenever necessary. To illustrate the irradiance
differences among the observed targets, we show a detail of FUSE spectra
in Figure \ref{figfuse}. The complete FUSE spectra are not shown for the
sake of clarity (see Guinan et al. 2003 for the identification of the
strongest features) and the region illustrated is a wavelength window
around two strong O~{\sc vi} emission lines. Again, a steep flux decrease
with increasing age is evident.

\subsection{UV: IUE \& HST} \label{sec:uv}

\begin{figure}[!ht]
\epsscale{0.5}
\plotone{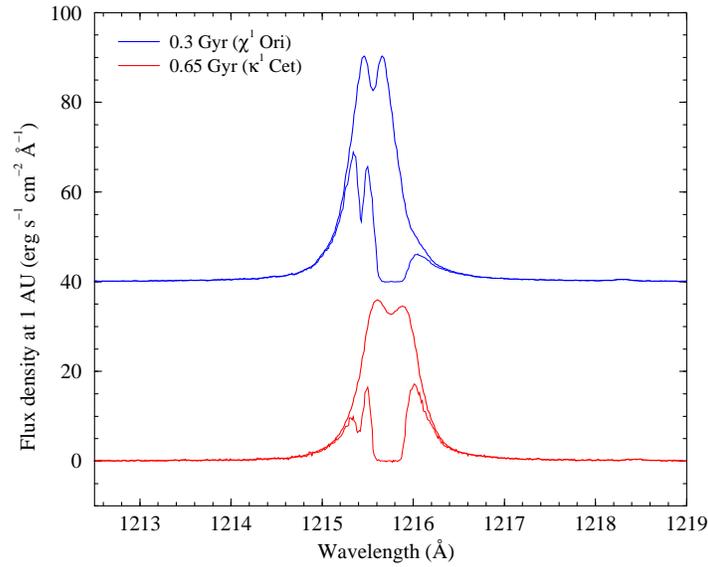}
\figcaption{Detail of the H~{\sc i} Lyman$\alpha$ line of two solar-type
stars of different ages. The thin lines show the observed profiles from 
HST spectra while the thick lines depict the reconstructed line profiles
after correction for H~{\sc i} and D~{\sc i} ISM absorption. Irradiances 
are obtained by integrating the absorption-corrected profiles. The top 
spectrum has been shifted by 40 erg s$^{-1}$ cm$^{-2}$ \AA$^{-1}$ to avoid 
confusion.  
\label{fighst}}
\end{figure}

UV spectroscopic observations of two of the stars in Table \ref{tabtarg}
($\chi^1$ Ori and $\kappa^1$ Cet) were obtained with HST within program
\#8280. Thanks to the small aperture and the large spectral resolution
used (STIS/E140M echelle grating and $0\farcs2$$\times$$0\farcs2$
aperture), these observations are ideally suited to study the strong
H~{\sc i} Ly$\alpha$ emission line, besides other emission lines in the
UV. The estimation of the integrated flux of the H~{\sc i} Ly$\alpha$ line
requires careful correction for interstellar H~{\sc i} and D~{\sc i}
absorption, which is significant even for the low column densities of our
targets. Full details on this procedure and the datasets used can be found
in Wood et al. (2004). A comparison of the H~{\sc i} Ly$\alpha$ emission
features (with both the observed and rectified line profiles) for the two
targets observed is shown in Figure \ref{fighst}.

\begin{figure}[!ht]
\epsscale{0.5}
\plotone{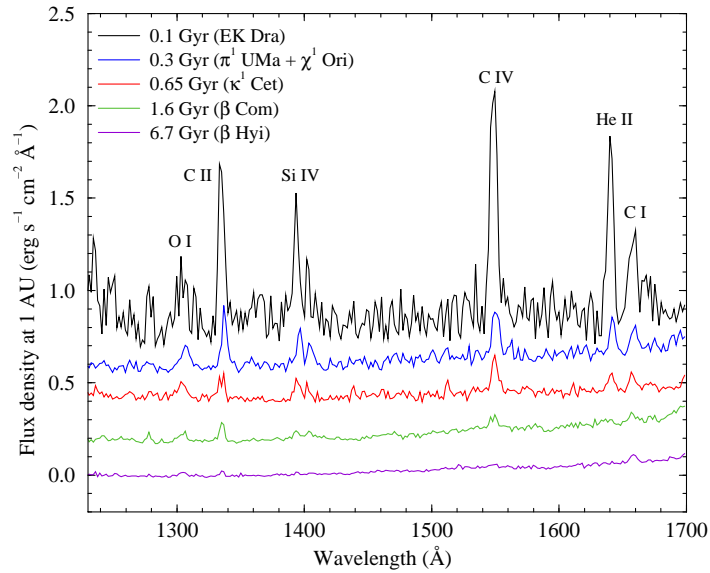}
\figcaption{UV irradiances of solar-type stars at different stages of 
evolution. Some relevant features are identified. The emission line strength 
is found to decrease from top to bottom (i.e., increasing age and rotation 
period). The spectra have been zeropoint-shifted using integer multiples of 
0.2 erg s$^{-1}$ cm$^{-2}$ \AA$^{-1}$ to avoid confusion. When compared with 
the EUVE and FUSE spectra, the emission lines seem broader here because of 
the lower spectral resolution of the observations. Note the onset of some 
weak photospheric (continuum) flux above 1500~\AA. 
\label{figiue}}
\end{figure}

The UV irradiances of the target stars up to 1700~\AA\ were completed with
IUE short wavelength camera (SWP) low-dispersion observations (e.g., Kondo
et al. 1989). Flux-calibrated data files are available from the MAST
archive following the NEWSIPS calibration pipeline (Nichols \& Linsky
1996). However, further correction to the fluxes was applied in accordance
with the investigation of Massa \& Fitzpatrick (2000). The bulk of the
flux in the UV region shortwards of 1700~\AA\ is in the form of emission
lines, with negligibly small photospheric continuum contribution. The
comparison in Figure \ref{figiue} depicts IUE spectra of the observed
stars, where the decrease of emission line fluxes with age is apparent.
The strongest features in the IUE range are labeled in Figure \ref{figiue}
and correspond to: O~{\sc i} $\lambda$1304 (triplet), C~{\sc ii}
$\lambda$1335 (triplet\footnote{One of the components is very weak and
separated only 0.04~\AA\ from a strong component.}), Si~{\sc iv}
$\lambda$1400 (doublet), C~{\sc iv} $\lambda$1550 (doublet), He~{\sc ii}
$\lambda$1640 (multiplet), and C~{\sc i} $\lambda$1657 (multiplet). As
mentioned above an important issue for some of the emission lines is the
correction of ISM absorption. Given the low electron density in the ISM
(e.g., Spitzer \& Fitzpatrick 1993; Wood \& Linsky 1997), only transitions
involving ground levels are expected to experience noticeable ISM
absorption.  All of the emission features listed above have at least one
ground level transition component except for He~{\sc ii} $\lambda$1640. In
the case of the lines of Si~{\sc iv}, C~{\sc iv}, and C~{\sc i}, the very
low abundance of these species in the ISM (largely dominated by C~{\sc ii}
and Si~{\sc ii}; see Wood et al. 2002b) implies negligible absorption in
the lines of sight of our targets, rendering any correction unnecessary.
In the case of the O~{\sc i} $\lambda$1304 and C~{\sc ii} $\lambda$1335
triplets, one of the components has a ground state transition that is
prone to ISM absorption, which we corrected as explained below.

To complement the IUE UV data we made use of the HST echelle spectra
described above for $\chi^1$ Ori and $\kappa^1$ Cet, which cover a
wavelength interval from 1150 to 1700~\AA. These higher resolution data
served the double purpose of cross-checking the integrated fluxes and
carrying out a direct correction for ISM absorption. Inspection of the
line profiles indicate that ISM absorption is present in the O~{\sc i}
$\lambda$1302.17 and C~{\sc ii} $\lambda$1334.53 components. These account
for about 30\% and 40\% of the total flux of the O~{\sc i} and C~{\sc ii}
triplets, respectively. We carried out a reconstruction of the line
profiles assuming a gaussian functional form with a superimposed
absorption (a total of six free parameters). For $\kappa^1$ Cet, the
difference in radial velocity between the star and the ISM components is
of only $\sim$5 km s$^{-1}$, thus implying significant absorption. In this
case, the corrections to the total O~{\sc i} $\lambda$1304 and C~{\sc ii}
$\lambda$1335 fluxes were of 14\% and 20\% respectively. The resulting
profile reconstruction parameters were found to agree very well with the
ISM properties in the $\kappa^1$ Cet line of sight (Wood et al.  2004).
For $\chi^1$ Ori, the radial velocity difference between the star and the
absorbing ISM is $\sim$36 km s$^{-1}$, which makes the flux correction
negligible ($<$2\%). Approximate corrections for the rest of the stars
were computed from the results on $\chi^1$ Ori and $\kappa^1$ Cet and the
properties of the ISM absorbing components in Wood et al. (2004) -- using
the line of sight proxies discussed in \S \ref{sec:targ}. The resulting
corrections to the total O~{\sc i} $\lambda$1304 and C~{\sc ii}
$\lambda$1335 fluxes were in the range 2--17\%.

\subsection{Solar irradiance} \label{sec:solar}

Finally, we shall describe the mean solar irradiance spectrum used in our
comparisons. As a consequence of an increasing number of sounding rocket
experiments and space missions, the knowledge of the Sun's high-energy
emissions has improved considerably in recent times. In our investigation
we have made use of the detailed irradiances constructed by Woods et al.
(1998 and references therein) from measurements made by sounding rockets
and the UARS SOLSTICE mission. The data products available\footnote{See
http://lasp.colorado.edu/rocket/rocket\_results.html} include
low-resolution (10~\AA) spectra and line-integrated fluxes for 1992, 1993,
and 1994, and a high-resolution (1~\AA) spectrum for 1994. All these years
correspond to Solar Cycle 22, with 1992 and 1993 representing moderate
solar activity levels, and 1994 representing the quiet Sun. We used data
from 1993, which corresponds to mid cycle, as representative of the Sun at
the ``average'' activity level. The rocket and UARS SOLSTICE observations
cover a wavelength interval from 20 to 2000~\AA. Because of the relatively
low resolution of the observations ($\sim$2~\AA), we used the higher
resolution solar data from the SMM/UVSP
atlas\footnote{ftp://umbra.nascom.nasa.gov/pub/uv\_atlases/} and the
high-resolution spectrum of $\alpha$ Cen A of Pagano et al. (2004) to
check for possible contamination by blends in the interval
$\lambda>1300$~\AA. Complementary soft X-ray irradiance measurements from
the SNOE experiment (Bailey et al. 2000) were also used in the interval
20--200~\AA.  To complete the necessary short wavelength data (hard X
rays), we made use of the inferred solar flux in the ASCA wavelength
interval (1--20~\AA) by G\"udel et al. (1997a). The resulting energy
distribution is in general good agreement with the measurements made by
the GOES 10 \& 12 satellites in the narrower 1--8~\AA\ range.

\section{Measurement and estimation of irradiances} \label{sec:irrad}

One of the ultimate goals of the ``Sun in Time'' project is to produce a
set of data products describing the high-energy irradiance evolution of
the Sun and solar-type stars across the main sequence. These fluxes can
then be used as an input to model Solar System planet and exoplanet
atmospheres and study their variations over time.  To present the
observational data described in \S \ref{sec:data} we distinguish two
separate components: Integrated fluxes in wide wavelength intervals
(roughly defined by the intervals covered by different missions) and
emission fluxes of the strongest features in the high-energy spectrum.
Data for five representative age stages are given: 100 Myr (EK Dra), 300
Myr ($\pi^1$ UMa and $\chi^1$ Ori), 650 Myr ($\kappa^1$ Cet), 1.6 Gyr
($\beta$ Com), 4.56 Gyr (Sun), and 6.7 Gyr ($\beta$ Hyi). In the case of
300 Myr, with two stellar proxies, we scaled the fluxes and computed
averages whenever data for the two stars were available.

\subsection{Integrated irradiances}

The intervals considered are: 1--20~\AA\ (ASCA), 20--100~\AA\ (ROSAT),
100--360~\AA\ (EUVE), and 920--1180~\AA\ (FUSE). Above 1180~\AA\ we
provide only the fluxes of the strongest features because of the
difficulty in achieving a reliable total integration (caused by the
increasing photospheric contribution). We will address this issue in a
forthcoming publication.  The total fluxes of the observed spectra were
corrected to a distance of 1 AU and scaled (using a simple radius-squared
relationship) to the expected radius of the Sun at the star's age (in
Table \ref{tabtarg}). The resulting stellar fluxes are presented in Table
\ref{tabfltot} (values not in parentheses in the first five rows).

\begin{figure}[!ht]
\epsscale{1.0}
\plotone{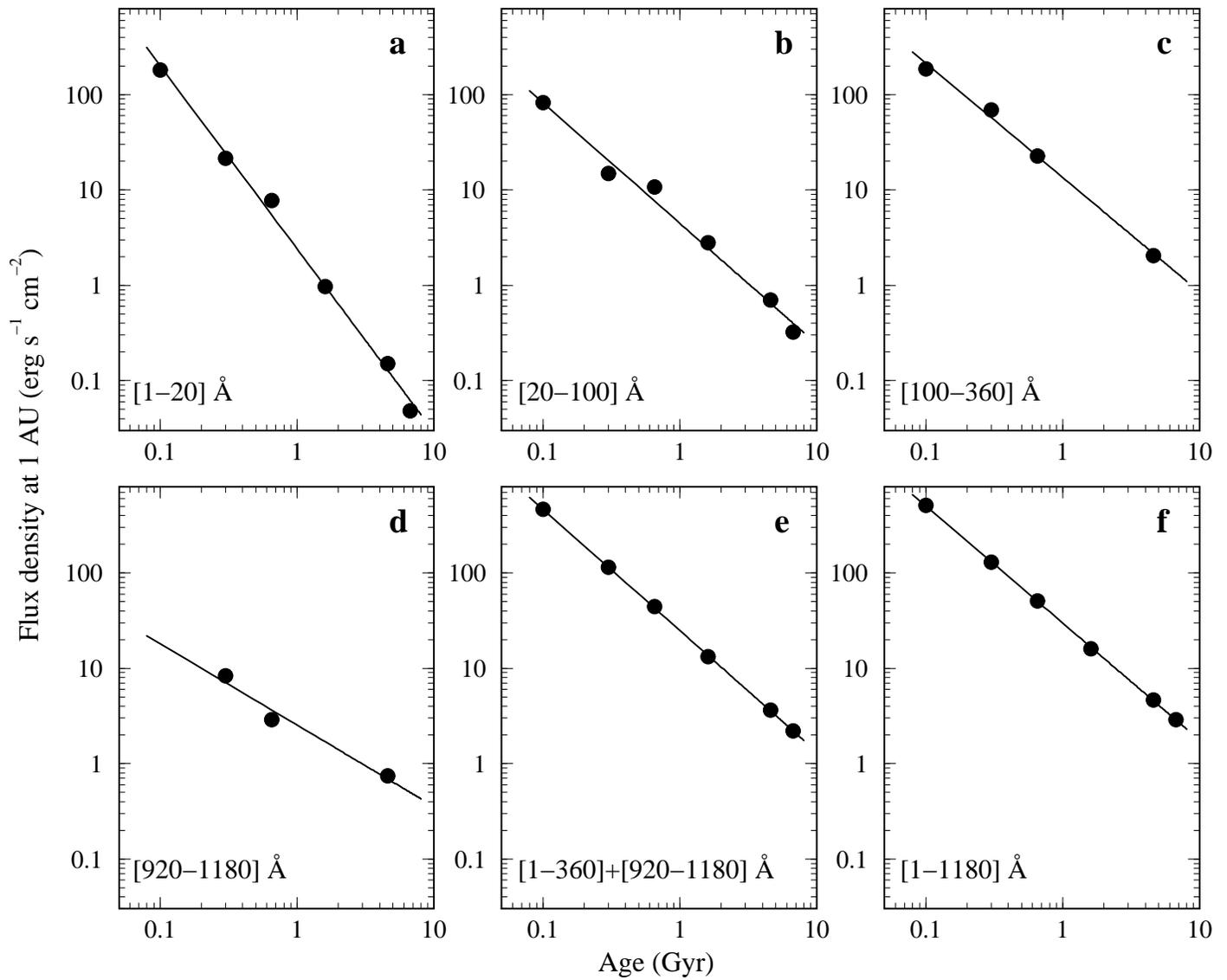}
\figcaption{Power-law fits to the integrated irradiances in Table
\ref{tabfltot} for different wavelength intervals. The parameters of the
resulting best-fitting relationships are given in Table \ref{tabslop}. 
\label{figpanels}}
\end{figure}

The measured fluxes show excellent correlations with the stellar ages (or,
equivalently, with the rotation periods). The analysis reveals that the
stellar fluxes can be very well approximated by power-law relationships,
as illustrated in Fig. \ref{figpanels}a-d. The parameters of the power-law
fits are given in Table \ref{tabslop}. Interestingly, the slopes of the
best-fitting relationships are seen to decrease monotonically from the
X-rays to the UV (i.e., decreasing energy or increasing wavelength).  
Thus, emissions associated with hotter plasmas are found to diminish more
rapidly and the overall plasma cools down as the stars spin down with age.
This behavior was already reported by earlier studies such as those by
Ayres et al. (1981) and Ayres (1999). Ayres (1997) gives also
relationships for the relative flux variations at different wavelengths.

As noted previously, the available observations are not complete because
of two chief reasons. First, there is a gap between 360 and 920~\AA\ for
stars other than the Sun, and second, measurements could not be obtained
for all targets in all wavelength intervals. The following targets have
observations missing in some wavelength intervals: EK Dra (FUSE), $\beta$
Com (EUVE, FUSE), $\beta$ Hyi (EUVE, FUSE). Note that, in the case of FUSE
observations, the lack of total flux values is caused by the impossibility
of calculating the flux contribution from the H Lyman lines because of
strong geocoronal contamination and saturated interstellar absorption (see
Guinan et al. 2003 for further explanation). In the case of EUVE, two of
the targets were not observed because their predicted fluxes were below
the detection limit of the instrument.

Having a (rough) estimate of the flux emitted by the stars in the
intervals not covered by observations is rather critical, especially if
the irradiances calculated in this work are to be used as input data in
planetary atmosphere modeling. The estimation of fluxes for stars with no
FUSE or EUVE observations is straightforward since the derived power laws
can be employed to make predictions. Using this procedure we calculated
the predicted fluxes in Table \ref{tabfltot} (values in single parentheses
in the first four rows) and a power-law fit to the integrated fluxes in
the wavelength interval covered by the observations (Fig.
\ref{figpanels}e). With regards to the 360--920~\AA\ interval, no
expectations exist of having any observational data, even in the
medium-term future. To circumvent the problem, there are at least three
possible alternatives: {\em 1)} Using empirical irradiances for the Sun
with some {\em ad hoc} scalings to account for the various activity levels
of our targets; {\em 2)} modelling from EUVE and UV by extrapolation under
the assumption that the lines and the continuum are from the same plasma
components; or {\em 3)} inferring the total integrated flux in the
interval by comparison with the flux evolution in other wavelength ranges.
We took the third approach chiefly because the accuracy we require is not
very high (some 10--20\% is sufficient as discussed below). In this way we
avoid the use of solar data not suitable to the very high activity level
of some of our targets. Also, the second method involves extrapolation
from an incomplete range of plasma temperatures since the observations
only cover coronal temperatures ($\sim$MK) and the temperatures of a few
FUV lines ($\sim10^5$ K). Given the power-law slopes in Table
\ref{tabslop} for different wavelength regions, a value of $-$1.0 seems to
be a good compromise in the 360--920~\AA\ interval. Thus, we calculated
the flux predictions in Table \ref{tabfltot} (values in double
parentheses) from the observed solar flux in this interval and the
inferred power-law relationship.  This crude interpolation could be flawed
if strong emission lines were present in the wavelength interval. This
does not seem to be the case when inspecting spectra from the Sun -- such
as the SOLSTICE data (see \S \ref{sec:solar}) or the SOHO/SUMER spectral
atlas (Curdt et al. 2001) -- and from Procyon (Drake et al. 1995), but
caution should be exercised when using the tabulated fluxes.

\begin{figure}[!ht]
\epsscale{1.0}
\plotone{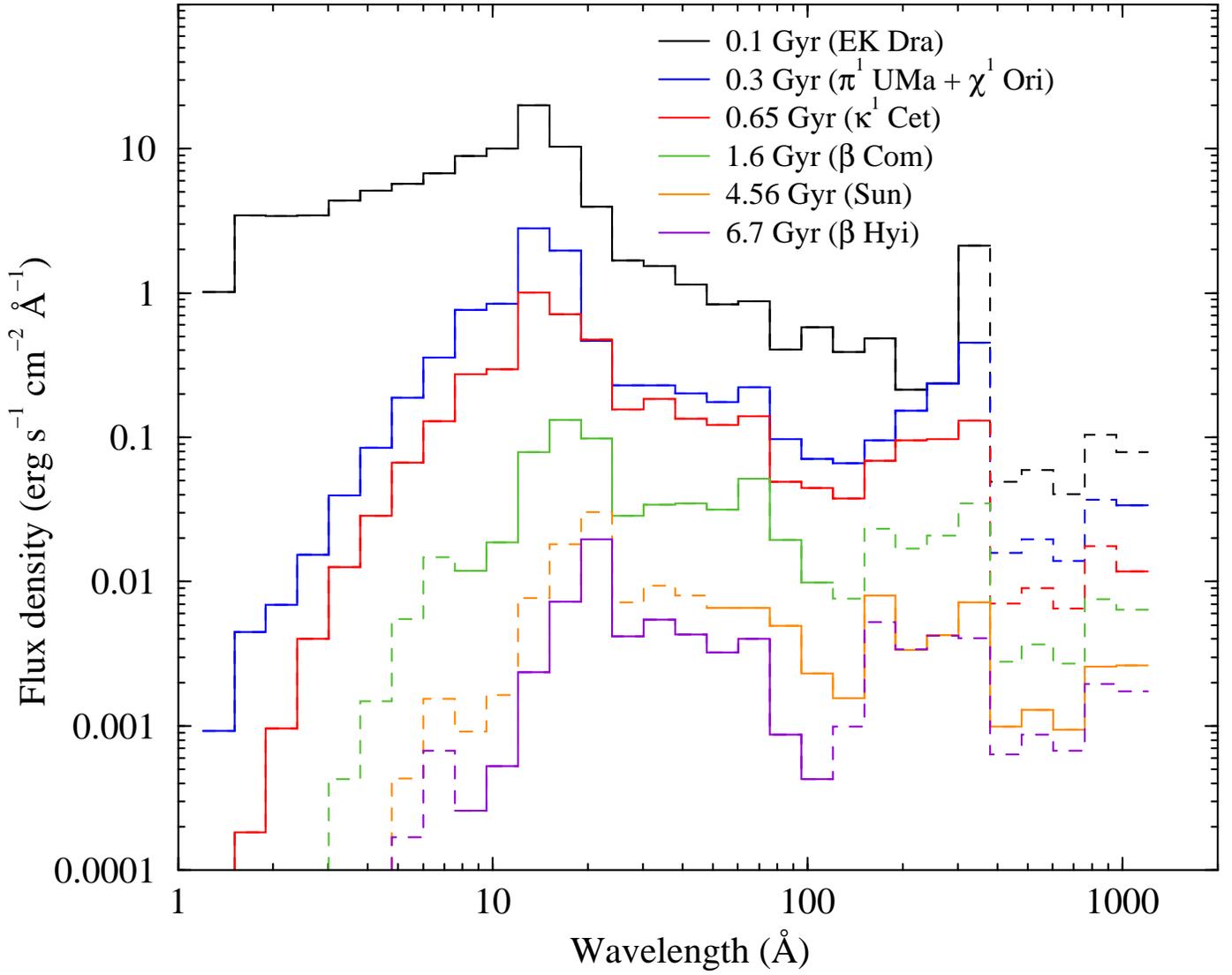}
\figcaption{Full spectral energy distribution of the solar-type stars at
different stages of the main sequence evolution. The solid lines represent 
measured fluxes while the dotted lines are fluxes calculated by interpolation 
using a power-law relationship. 
\label{figbins}}
\end{figure}

\begin{figure}[!ht]
\epsscale{1.0}
\plotone{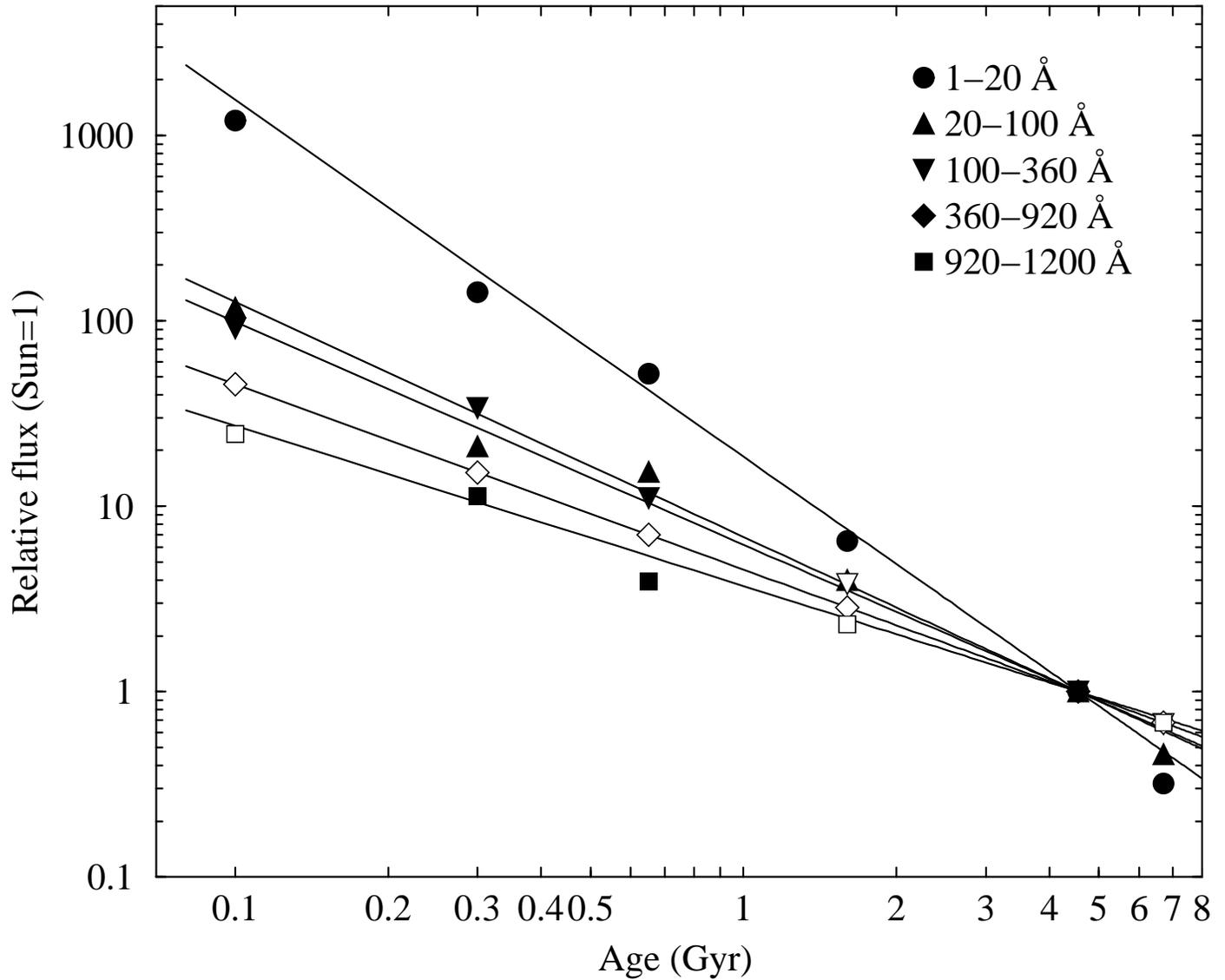}
\figcaption{Solar-normalized fluxes vs. age for different stages of the 
evolution of solar-type stars. Plotted here are the measurements for 
different wavelength intervals (filled symbols) and the corresponding 
fits using power-law relationships with the slopes in Table \ref{tabslop}. 
Represented with empty symbols are the inferred fluxes for those intervals 
with no available observations (values in parentheses in Table 
\ref{tabfltot}). 
\label{figfluxrel}}
\end{figure}

From the estimations described above, we were able to compute total
irradiances in the interval 1--1180~\AA\ as given in the last row of Table
\ref{tabfltot}. Also, in Fig. \ref{figpanels}f we plot the total
integrated fluxes and a power law fit that is found to yield an excellent
fit to the data. An illustration of the spectral energy distribution of
the targets is provided in Fig. \ref{figbins}, where the solid lines
represent the observed data and the dotted lines have been calculated via
power-law interpolation.  In the complete wavelength interval, which is
frequently used in aeronomy calculations, the flux ($F$) as a function of
stellar age ($\tau$) is accurately reproduced by the expression (the upper
wavelength limit can be extended from 1180 to 1200~\AA\ because there are
no additional relevant emission lines):
\begin{equation}
F = 29.7 \cdot \left[\tau \mbox{(Gyr)}\right]^{-1.23} \mbox{erg s$^{-1}$ cm$^{-2}$};
\hspace{1cm} \mbox{1 \AA} < \lambda < \mbox{1200 \AA}
\end{equation}
Also illustrative is the plot in Fig. \ref{figfluxrel}, which represents
the stellar fluxes normalized to the current solar values as a function of
age. The steeper decrease of the higher energy emissions is evident in
this plot. Also note that the emissions of the youngest stars in all
wavelength intervals are orders of magnitude larger the current solar
flux.

\subsection{Line fluxes} \label{sec:linflu}

Line integrated fluxes were measured for the strongest features of the
high-energy spectrum whenever possible (i.e., with sufficiently high
instrumental resolution). As explained above, the observed fluxes were
corrected for interstellar absorption (if necessary), scaled to a distance
of 1 AU and to the expected radius of the Sun at the star's age. The
resulting line fluxes are presented in Table \ref{tabflin}. Note that, as
mentioned in \S \ref{sec:solar}, the fluxes of the Sun for lines with
$\lambda>1300$~\AA\ were checked for blends with high-resolution spectra.  
We found any contamination to be below a negligible 10\% and no further
action was taken.

Similarly to the total integrated fluxes, the line fluxes in Table
\ref{tabflin} are also observed to follow well-defined power law
relationships as a function of stellar age, with the slopes given in Table
\ref{tabslin}. Also listed in this table are the characteristic ion
formation temperatures, which can be regarded as the typical temperatures
of the plasma where the line emissions are originating (Arnaud \&
Rothenflug 1985). Similarly to the integrated emissions, the flux decrease
with age becomes more pronounced with higher formation temperatures (or,
generally, shorter wavelengths).

Worth noting here is the powerful H~{\sc i} Lyman $\alpha$ emission
feature. As can be seen by comparing Tables \ref{tabfltot} and
\ref{tabflin}, Ly$\alpha$ is a very significant contributor to the
short-wave emission in the Sun and solar-type stars. This sole emission
line produces a large fraction of the total flux between 1 and 1700~\AA:
from about 20\% at 100 Myr up to over 50\% for the current Sun. The
observational data on Ly$\alpha$ irradiances of solar-type stars are still
scarce, with only two targets measured thus far. The preliminary results
presented here yield the following expression:
\begin{equation}
F = 19.2 \cdot \left[\tau \mbox{(Gyr)}\right]^{-0.72} \mbox{erg s$^{-1}$ cm$^{-2}$};
\hspace{1cm} \mbox{H~{\sc i} Lyman $\alpha$}
\end{equation}
Note that the formation temperature of this line is about 10 kK and so the
power-law slope should be less steep but similar to that in the
920--1180~\AA\ range, exactly as found (cf. Table \ref{tabslop}). For
comparison, the results of Wood et al. (2004) (i.e., $F(\mbox{Ly}\alpha)\propto
P_{rot}^{-1.09\pm0.08}$), when combined with the age--rotation period
relationship of Ayres (1997) (i.e., $P_{rot}\propto \tau^{-0.6\pm0.1}$), yield
$F(\mbox{Ly}\alpha)\propto \tau^{-0.65\pm0.12}$ for F and G dwarfs.
It is worth noting that this slope is in remarkably good agreement with
the value we find.

Some studies have also revealed correlations between certain line fluxes
and the overall high-energy flux. This is the case of Bruner \& McWhirter
(1988), who reported a tight correlation between the C~{\sc iv}
$\lambda$1550 flux and the total radiated power by solar active regions.
We have carried out a similar comparison between the C~{\sc iv}
$\lambda$1550 flux and the total integrated flux in the 1--1200~\AA\
interval (cf. table 7 and figure 8a of Bruner \& McWhirter 1988). The
power-law fit has a remarkable correlation coefficient of $0.998$ and
yields a slope of $1.12\pm0.04$ (i.e., $F_{\rm tot}\propto F_{\rm C
IV}^{1.12}$), in good agreement with the value $1.08$ obtained by Bruner
\& McWhirter (the somewhat steeper slope in our case is likely due to the
non-inclusion of the UV flux, which would flatten the relationship). This
is a strong argument in favor that the emissions of our targets arise from
active regions that are similar in nature to those of the Sun.

\section{Discussion}

The comprehensive investigation presented here unequivocally demonstrates
that the Sun has experienced a strong decrease of its high-energy
emissions over the course of its main sequence evolution. Quantitatively,
the results indicate that the the solar high-energy flux in the interval
1--1200~\AA\ was about 2.5 times the present value 2.5 Gyr ago and about 6
times the present value about 3.5 Gyr ago. Also, the 100 Myr ZAMS Sun
should have had high-energy emissions some 100 times larger than presently
in this wavelength interval. The great diminishing of the solar
high-energy flux with time is vividly illustrated by the following fact:
EK Dra's flux in the sole C~{\sc iii} $\lambda$977 emission line is larger
than the entire integrated current solar irradiance below 1200~\AA.

The results also show that an important contributor to the high-energy
emissions of solar-type stars is the strong H~{\sc i} Ly$\alpha$ feature.  
This statement remains true throughout the lifetime of the stars, although
the relative fraction of Ly$\alpha$ photons with respect to the
high-energy emissions increases with the age of the star. This
investigation shows that the Ly$\alpha$ flux 2.5 Gyr and 3.5 Gyr ago was
larger than today by factors 1.8 and 2.9, respectively. Again, the 100 Myr
old Sun was much more active, with an expected Ly$\alpha$ flux some 15
times larger than presently.

Note that the stellar sample we used covers the solar irradiance evolution
from an age of about 100 Myr after its arrival on the main sequence. There
are numerous indications that the Sun was even more active during the T
Tauri and the early post-ZAMS stages (e.g., Simon et al. 1985, and
references therein). Studies indicate that the X-ray luminosity of
solar-type stars reaches a saturation level ($\log L_{\rm X}/L_{\rm
bol}\approx -3$) at a rotation period of about 1.5 d (Pizzolato et al.
2003). The X-ray emission of the youngest, most active solar type stars
can be up to 2-3 times higher than the flux of our youngest solar proxy,
EK Dra. This is observed in stars of clusters such as $\alpha$ Per or IC
2391, with ages around 50 Myr. Analogous, saturation effects are expected
for the emissions in the EUV and UV ranges. The evolutionary stages
younger than 100 Myr have strong significance on stellar evolution, dynamo
theory, but also on the ionization of the accretion disk, hence on the
planetary formation, and on the astrochemistry. However, these early
stages do not bear special relevance to studies related to planetary
atmospheres and environments since planets were still forming in the
protoplanetary nebula (Chambers \& Wetherill 1998; Lissauer 1993). Even
somewhat later in the evolution (up to $\sim$500 Myr), the influence of
the strong solar irradiance may have been overwhelmed by the heavy
bombardment period in the inner Solar System (Sleep et al. 1989) as well
as planetary meltdowns and volcanism.

\subsection{Previous studies}

Similar previous works on the same subject were published by Zahnle \&
Walker (1982) and Ayres (1997). The study of Zahnle \& Walker (1982)
focused on the evolution of solar ultraviolet emissions and was triggered
by the early discoveries of high XUV luminosities of young late-type stars
made by the IUE and Einstein satellites.  The authors used T Tauri
(pre-main sequence) stars and the current solar flux values to interpolate
a flux evolution law for different wavelength intervals assuming a
$t^{-1/2}$ scaling law for the rotational velocity. The results are in
reasonable agreement with ours, in spite of the fact that Zahnle \& Walker
did not use a true solar analog sample and employed just two fiducial
points for the interpolation. Ayres (1997) carried out a very detailed
study of the solar high-energy flux, including a photoionization model for
four species (H, O, O$_2$, N$_2$) mostly focused on its application to the
primitive Martian atmosphere. The author used a combination of empirical
data (EUVE) with global scalings of the solar spectrum using power-laws
with slopes depending on the typical temperature formations of the studied
ion species. In spite of the different approach, the results by Ayres are
in general good agreement with ours. Perhaps the power-law slopes reported
by Ayres are slightly smaller, yet still compatible with those presented
here.

\subsection{Uncertainties, cycles, and variability}

We have presented in \S \ref{sec:irrad} the results of our XUV
observations, but no error estimates have been discussed yet for the flux
values. There are four chief sources of uncertainty in the stellar fluxes
provided (if we neglect the errors associated with the radius and ISM
absorption corrections): the measurement errors, the intrinsic variability
of the emission, and the scatter associated to the selection of the
stellar proxies (i.e., we are assigning a single flux to a mass or spectral
type interval). In turn, the error of the measurement has a contribution
from the flux integration itself (photon noise) and from the calibration
of the detector. The numerical error of the flux integration follows from
the propagated error of the the flux uncertainty in each wavelength bin.

We have carried out the necessary calculations and find that the
observations were made with sufficiently long integration times so that
the photon noise contributes to an uncertainty of less than 5\% on the
measured fluxes. The exceptions to this are the integrated FUV fluxes,
which have strong geocoronal contamination and had to be inferred from
alternative methods (see Guinan et al. 2003) resulting in uncertainties of
20--40\%, and the ROSAT X-ray flux of $\beta$ Hyi, with an uncertainty of
about 10\% due to the low count rate. The estimation of the calibration
errors is quite often not straightforward. From the comparison between
different X-ray missions and the absolute effective area calculations one
deduces an absolute calibration uncertainty of the order of 10--20\% for
both ASCA SIS and ROSAT PSPC detectors (see documentation in
http://heasarc.gsfc.nasa.gov/). In the case of EUVE, Bowyer et al. (1996)
report that the effective area calibration of each band is believed to be
accurate to within 20\%. For the FUSE LWRS LiF and SiC detectors, the
absolute calibration of the fluxes has an uncertainty below 10\% according
to the documentation in the FUSE
website\footnote{http://fuse.pha.jhu.edu/analysis/calfuse\_wp0.html}.
Detailed documentation is available on the absolute photometric accuracy
of HST STIS/MAMA echelle observations (as those used here) and the
expected value is around 8\% (see Space Telescope Imaging Spectrograph
Instrument Handbook at http://www.stsci.edu/hst/stis/documents and also
Bohlin 1998).  Finally, comparisons indicate that, after correction of
systematic effects, the absolute calibration of IUE is accurate to within
3\% (Massa \& Fitzpatrick 2000).

Summarizing, we estimate that the measurement errors (including both
photon noise and absolute calibration uncertainty) of the fluxes in Tables
\ref{tabfltot} and \ref{tabflin} decrease from about 10--15\% in the X-ray
domain to about 5-8\% in the UV.

As mentioned above, there are additional sources of uncertainty caused by
the intrinsic variability of the emission and by the differences between
stars in the studied spectral range. Recall that the ultimate goal of this
study is to obtain XUV fluxes for the Sun and solar-type stars over their
(magnetic) evolutionary histories using stellar proxies. We are interested
in obtaining the ``average'' XUV emissions that are characteristic of
stars in a specific mass or spectral type window (G0--5). With the
available observations, there is very little we can say about the scatter
caused by using stars that are slightly more massive than the Sun (see
Table \ref{tabtarg}) as proxies. However, there are two competing effects
that may cancel each other to a certain extent. On the one hand, more
massive stars have larger surface areas and thus larger integrated
emissions but, on the other hand, they also have shallower convective
zones and a somewhat weaker dynamo.  We may speculate that, at first
order, the emissions of stars within a small interval in masses are very
similar.

But certainly, the main contributor to the uncertainty of the fluxes in
Tables \ref{tabfltot} and \ref{tabflin} is caused by the intrinsic
variability. It is well known that stellar magnetic activity is
characterized by short- and mid-term variations over timescales of hours,
days, months, and years. It is beyond the scope of this paper to analyze
in detail the flux variations of all our solar proxies over these
timescales because that would imply a very large observational effort. In
most cases, the available observations represent just a ``snapshot'' of
the flux emissions without any reference to the ``average'' value and its
scatter. We may, however, make an educated guess at the amplitude of the
variations over timescales of years. In the Sun, besides flares, which are
discussed below, the relevant source of mid-term variability is the 11-yr
activity cycle (see Lean 1997 for a complete review). In spite of the lack
of accurate solar XUV variability measurements yet, the available data
indicate solar maximum vs. minimum (i.e., peak to peak) flux ratios of
10--20 for X-rays (10--100~\AA) decreasing to ratios of about 2 at
600~\AA\ and 1.2 at 1500~\AA\ (Hinteregger 1981; Rottman 1988; Lean 1997).
In the case of the H~{\sc i} Lyman $\alpha$ emission line, Woods et al.  
(2000) carry out a detailed study over different timescales and report a
solar-cycle peak-to-peak flux ratio of 1.5. The variability factor shows a
positive correlation with the temperature of the associated emitting
plasma. Thus, for specific lines, the flux variations are a function of
their formation temperatures (see Ayres 1997).

Solar-like stars have also been observed to exhibit activity cycles
similar in length to that of the Sun (e.g., Baliunas \& Vaughan 1985). For
example, there is some yet inconclusive evidence for a 10--12 yr X-ray
activity cycle in the case of EK Dra (G\"udel et al. 2003) and a cycle of
similar length in $\beta$ Hyi (Guinan, unpub). Interestingly, however, the
flux maximum vs. minimum ratio for EK Dra in the ROSAT band (i.e.,
$\approx$6--120~\AA) is about 2.5, or some 4 times smaller than the Sun's
value (e.g., Hempelmann et al. 1996; Micela \& Marino 2003). This may just
be a consequence of the smaller contrast between maximum and minimum when
the stars have high activity levels (i.e., surface active regions) at all
times.  Indication for possible long-term modulation depths can be
obtained by observing cluster stars repeatedly, separated by several
years. Such studies have, for example, been performed for the Pleiades
cluster that contains many stars somewhat similar to EK Dra, or for the
Hyades, similar in age to $\kappa^1$ Cet. These observing programs
typically reported variations of no more than a factor of two for most
stars, see, for example, Gagn\'e et al. (1995), Micela et al. (1996), and
Marino et al. (2003) for the Pleiades and Stern et al. (1994) and Stern et
al. (1995) for the the Hyades.

The flux intrinsic variability with the activity cycle of our solar
proxies will be the subject of a forthcoming study when the time baseline
of the data permits a detailed investigation. At this point, taking into
account all the available information, we estimate total uncertainties for
the power-law slopes in Table \ref{tabslop} no larger than 0.1.

A further source of short-term variability are stellar flares. These are
important to certain applications of our irradiance data because the
amount of energy released in a single event can be significant.
Observations of some of the solar proxies in our stellar sample indicate
that flare events in young solar proxies such as EK Dra are frequent
($\sim$3--4 major flares per day) and up to 100 times more powerful than
observed for the present Sun (Audard et al. 1999). X-ray flaring has also
been observed on $\pi^1$~UMa and a large X-ray flare (ten-fold X-ray
enhancement) was recorded by the EXOSAT satellite during January 1984
(Landini et al. 1986). Another of our targets, $\kappa^1$ Cet, experienced
a flare that was recorded spectroscopically in the visible by Robinson \&
Bopp (1987). For an up-to-date study of X-ray flaring (which includes new
observations of some of our targets) the reader is referred to the recent
paper by Telleschi et al. (2005). Also, we plan in the near future to use
recently acquired time-tagged spectra with FUSE to address the evolution
of the flare rates and energetics of solar-type stars over their
lifetimes. We stress that the fluxes given in this paper are clear of
major flare events and should constitute a faithful representation of the
quiescent emission of the targets.

\subsection{Particle fluxes}

With enhanced high-energy emissions and frequent flares, young solar-type
stars are also expected to have more powerful particle winds. Evidence
from lunar and meteoritic fossil record agrees with this extrapolation and
suggests that the Sun had a stronger wind in the past (e.g., Newkirk
1980). Similar conclusions were drawn by Lammer et al. (2000) from the
study of the $^{15}$N/$^{14}$N isotope ratio in the atmosphere of Titan.
The indirect evidence of an enhanced particle flux during the first 500
Myr of its life would be more compelling if a direct detection of the wind
of a solar-type star was attained.  Attempts made by Gaidos et al. (2000)
to detect the winds of $\pi^1$ UMa, $\kappa^1$ Cet, and $\beta$ Com from
their radio emissions yielded negative results. Also unsuccessful was the
search for blue-shifted absorption in coronal lines carried out by Ayres
et al. (2001).

Although winds of solar-like stars have not yet been detected directly,
Wood et al. (2001, 2002a) devised a method to infer their characteristics
from observations of the interaction between the fully ionized coronal
winds and the partially ionized local ISM. Modeling of the associated
absorption features, which are formed in the ``astrospheres,'' has
provided the first empirical estimates of coronal mass-loss rates for G--K
main-sequence stars. From the small sample where the astrospheres can be
observed, the mass loss rates appear to increase with stellar activity.
Using simple relationships involving rotational velocities and X-ray
fluxes, Wood et al. (2002a) suggest that the mass loss rate of the Sun has
decreased following a power law proportional to $t^{-2}$, which implies
that the wind of the active young Sun may have been around 1000 times more
massive than it is today. There are still a number of assumptions that
have to be proved before this result can be fully established, especially
the correlation between X-ray flux and mass loss rate as these originate
from physically distinct regions (closed and open fields, respectively),
but Wood et al.'s work is an important step forward.

With many of the solar wind characteristics still being unveiled today, it
is not surprising that our knowledge of the particle fluxes of solar-type
stars of different ages is at a very basic stage. Evidence from
independent sources indicates that the young Sun (and by extension, young
solar-type stars) had a wind significantly more intense than presently.
Also, the high frequency of large flares observed with EUVE by Audard et
al. (2000) in young Suns such as EK Dra and 47 Cas~B could indicate
explosive episodic releases of plasma generating non-thermal high-energy
particles. These would be like the coronal mass ejections observed on the
Sun but hundreds of times stronger and more frequent. Similarly to the Sun
today (Lewis \& Simnett 2000; Schrijver \& Zwann 2000), coronal mass
ejections could contribute significantly to the stellar wind. As we
discuss below, the solar wind plays an important role in the shaping and
evolution of planetary atmospheres and surfaces and thus it is an
important component when characterizing the magnetic activity evolution of
the Sun over its lifetime.

\section{Applications of the solar irradiance data}

\subsection{Thermal escape on exoplanets}

The high-energy irradiance evolution data presented in this paper are
intended to be an ingredient of studies related to the evolution of Solar
System planets and exoplanets. Meier (1991) gives the absorption profile
of the Earth's atmosphere and shows that most of the radiation with
$\lambda\la1700$~\AA\ is absorbed (and thus deposits its energy) in the
thermosphere, at an altitude above 90 km. Similar effects are found in
atmospheres with different compositions (e.g., within the Solar System)
since the absorption cross-sections of the high-energy photons by all the
atomic species present in the upper atmospheres are very large. Thus, XUV
ionizing radiation raises the temperature of the planetary thermospheres
and affects their vertical temperature profiles and energy transport
mechanisms. Obviously, planets around young solar-type stars, with XUV
fluxes 10--100 times stronger than today's Sun, suffer intense heating of
their upper atmospheres, that reach temperatures much above the current
value of 1000~K for the Earth. When the temperature of the thermosphere is
large, a significant fraction of the light constituents of the upper
atmosphere attain velocities above the escape value (and drag heavier
constituents away).

Thermal escape, although commonly neglected in the present Solar System,
may be important in planets around magnetically active stars. Thus, any
attempt to calculate the history of a planet's atmosphere around a
solar-type star needs to include the XUV energy evolution, since this
regulates the efficiency of evaporation processes. Calculations using
early data from the present work were carried out by Lammer et al.
(2003b), who showed that ``Hot Jupiters'' could lose significant fractions
of their hydrogen masses under intense XUV radiation. Non-thermal
mechanisms caused by ionosphere-stellar wind interactions also contribute
to this loss processes (Grie$\beta$meier et al. 2004). The tantalizing
results indicate that hydrogen-rich giant exoplanets may suffer rapid
evaporation under strong XUV radiation conditions. Once the entire
hydrogen envelope is lost, only the rocky planetary cores would remain,
thus representing a putative new class of planet. The confirmation of the
theoretical prediction of thermal escape comes from the observations of
Vidal-Madjar et al. (2003), who reported a large exospheric radius for the
transiting planet HD 209458 b (due to thermal expansion) and a loss rate
compatible with the estimates of Lammer et al. The consequences of this
enhanced thermal loss process could explain the apparent paucity of
exoplanets so far detected at very close orbital distances ($<0.05$ AU).  
Terrestrial-like planets could also be affected by the enhanced XUV
environment and lose a significant fraction of their lighter atmospheric
constituents.

\subsection{The Martian water inventory}

The ``Sun in Time'' data are also being used to study aspects related to
the evolution of Solar System planet atmospheres and surfaces. In
particular, the planet Mars has been especially vulnerable in the past to
the influence of the Sun's energy and particle emissions because of its
small mass and the lack of a protecting magnetic field. Lammer et al.
(2003a) and Terada et al. (2004) have studied the Martian water inventory
using reliable solar XUV and wind evolution laws and comprehensive models
for the loss processes of hydrogen and oxygen that include dissociative
recombination, ion pickup, sputtering, viscous processes in the planet's
ionosphere. The more recent work of Terada et al. uses a global hybrid
model to conclude that the loss of H$_2$O from Mars over the last 3.5 Gyr
is equivalent to a global Martian ocean with a depth of about 10.5 m. This
value is smaller than those reported by previous studies but could still
be slightly overestimated.

The two studies quoted also find that the sum of thermal and non-thermal
atmospheric loss rates of H and all non-thermal escape processes of O to
space are not compatible with a ratio of 2:1 (H to O) expected from the
atomic composition of water, and is currently close to about 20:1. Escape
to space cannot therefore be the only sink for oxygen on Mars. These
results suggest that the missing oxygen can be explained by the
incorporation into the Martian surface by chemical weathering processes
since the onset of intense oxidation about 2 Gyr ago. The oxygen
incorporation has also implications for the oxidant extinction depth,
which is an important parameter to determine required sampling depths on
Mars aimed at finding organic material. The oxidant extinction depth is
expected to lie in a range between 2 and 5 m for global mean values.

\subsection{Erosion of Mercury's surface}

The planet Mercury, because of its closeness to the Sun, has suffered
major exposure to the particle and XUV emissions during the early active
stages (see Guinan \& Ribas 2004). Mercury's core is large compared to
other terrestrial planets, extending out to over 60\% of its radius. One
of several hypotheses advanced to explain this anomaly is that strong,
dense winds and very high XUV fluxes of the young Sun (during the first
0.5--1 Gyr of its life) swept away its early atmosphere and much of its
outer mantle. Even today (with a much less active Sun) ground based
observations of heavy constituents like Na$^+$, K$^+$ and O$^+$ in
Mercury's present transient exosphere implicate a strong exosphere-surface
interaction related to the particle and radiation environment of the
nearby Sun (e.g., Cameron 1985). Lammer et al. (2002) have carried out
initial calculations that indicate that enhanced solar wind and XUV
emissions could be sufficient to explain the present relatively thin
mantle and relatively large iron core. If this hypothesis is correct,
young Mercury may have started out similar in size to the Earth but lost
much of its less dense mantle from radiation and particle interactions
(ion pick-up) with the young Sun.

\subsection{The paleo-climate of the Earth}

Finally, the young Sun's emissions may have also had an impact on the
early evolution of Earth's atmosphere. Besides heating the thermosphere
and altering the vertical temperature profile, the enhanced high-energy
flux can strongly influence the photochemistry and photoionization of the
early planetary atmospheres and also may play a role in the origin and
development of life on Earth as well as possibly on Mars. For example,
Canuto et al. (1982, 1983) discuss the photochemistry of O$_2$, O$_3$,
CO$_2$, H$_2$O, etc, in the presumed CO$_2$-rich early atmosphere of the
Earth. In this context, the Ly$\alpha$ flux plays an important role as it
is strong enough to penetrate the planetary exospheres into their
mesospheres, richer in molecules and susceptible to photochemical
reactions.

The ``Sun in Time'' data can also provide insights into the so-called
Faint Sun Paradox (see Guinan \& Ribas 2002). The paradox arises from the
fact that standard stellar evolutionary models show that the Zero-Age Main
Sequence Sun had a bolometric luminosity of $\sim$70\% of the present Sun.  
This should have led to a much cooler Earth in the past while geological
and fossil evidence indicate otherwise. A solution to the Faint Sun
Paradox proposed by Sagan \& Mullen (1972) was an increase of the
greenhouse effect for the early Earth. The gases that have been suggested
to account for this enhanced greenhouse effect are CO$_2$, NH$_3$ or
CH$_4$ (see, e.g., Rye et al.  1995; Sagan \& Chyba 1997; Pavlov et al.
2000).  Although the stronger XUV solar radiation cannot by itself explain
the Faint Sun Paradox (because it only accounts for an insignificant
percentage of the Sun's radiative output), the photoionization and
photodissociation reactions triggered could play a major role in what
greenhouse gases are available. For example, the enhanced
photodissociating FUV--UV radiation levels of the young Sun may have
influenced the abundances of ammonia and methane in the pre-biotic and
Archean planetary atmosphere some 2-4 Gyr ago. Similarly, the
photochemistry and abundance of O$_3$ is of great importance to study life
genesis on Earth.

In summary, any model of the paleoatmosphere of the Earth and other Solar
System planets needs to account for the stronger XUV and particle
radiation from the young Sun. If the XUV fluxes of the young Sun are
estimated by a simple scaling of the current values by a factor of 0.7 (in
accordance with the lower expected photospheric flux), this will represent
a severe underestimation of the radiation levels by orders of magnitude.

\section{Conclusions}

One of the primary goals of the ``Sun in Time'' program is to reconstruct
the spectral irradiance evolution of the Sun and, by extension, of
solar-type stars. To this end, a large number of multiwavelength (X-ray,
EUV, FUV, UV, visible) have been collected and analyzed. The observations,
secured with the ASCA, ROSAT, EUVE, FUSE, HST, and IUE satellites, cover
1~\AA\ (12~keV) to 1700~\AA, except for a gap between 360~\AA\ and
920~\AA, which is a region of very strong ISM absorption. Irradiance data
have are already available for five of the stars in our sample.

A detailed quantitative analysis reveals that the stellar fluxes can be
very well approximated by power-law relationships. Interestingly, the
slopes of the best-fitting relationships are seen to decrease
monotonically from the X-rays to the UV (i.e., decreasing energy or
increasing wavelength). Emissions associated with hotter plasmas diminish
more rapidly and the overall plasma cools down as the stars spin down with
age. The results from the ``Sun in Time'' program suggest that the coronal
X-ray--EUV emissions of the young main-sequence Sun were $\sim$100--1000
times stronger than those of the present Sun. Similarly, the transition
region and chromospheric FUV--UV emissions of the young Sun are expected
to be 20--60 and 10--20 times stronger, respectively, than presently. In
the entire XUV interval from 1 to 1200~\AA\, we find that the solar
high-energy flux was about 2.5 times the present value 2.5 Gyr ago and
about 6 times the present value about 3.5 Gyr ago (when life arose on
Earth). Also, preliminary estimates using spectra of two solar proxies
indicate that Ly$\alpha$ flux of the young Sun was also much stronger, by
up to a factor of 15. In addition to intense levels of dynamo generated
coronal and chromospheric XUV emissions, the young Sun and young solar
analogues are also expected to have stronger and more frequent flares and
to have stronger (more massive) stellar winds.

In summary, compelling observational evidence indicates that the Sun
underwent a much more active phase in the past. The enhanced activity
revealed itself in the form of strong high-energy emissions, frequent
flares, and a powerful stellar wind. Such energy and particle environment
certainly had an impact on the genesis and evolution of Solar System
planets and planetary atmospheres.

Besides completing remaining gaps in the data (e.g., Ly$\alpha$
irradiances; somewhat complicated by the demise of the STIS spectrograph
on HST) and better characterizing the flare statistics and wind
properties, future work will be directed in two main directions. We plan
to extend our study to longer wavelengths between 1700 and 3000~\AA.  
Here the photospheric emissions begin to dominate over those of the
chromosphere and a lot more care has to be taken in adequately normalizing
the flux. Also, because the emission differences between young and old
solar-type stars are expected to be smaller ($\sim$10-30\%). The UV
portion of the spectrum (UVA, UVB, and UVC) is of great importance as it
drives the majority of the photochemical reactions and may influence the
generation and destruction of some chemical compounds, e.g., NO$_{\rm x}$,
of importance to life.

Because lower-mass stars are especially common and hence may host
habitable planets, progress in this direction has started by expanding the
``Sun in Time'' program to time sequences of the high-energy emissions,
wind, and flare activity of low-mass K and M stars. These stars are far
more numerous than the solar-type stars, have long main sequence lifetimes
and are, in principle, prime targets for terrestrial planet searches.  
Because of the low luminosities, their ``habitable zones'' (see Kasting et
al. 1993) can be quite close to the host stars. Low-mass stars have deeper
outer convective zones (where the magnetic dynamo operates) than Sun-like
stars and thus possess very efficient magnetic dynamos. The expected
enhanced XUV radiation environment should play a major role in the
development of the atmospheres and ultimately of life on planets located
in their habitable zones.

\acknowledgements

We gratefully acknowledge B. E. Wood for his help with the analysis of the
Lyman $\alpha$ observations. We are also grateful to H. Lammer and F.
Selsis for their support to this project and for their interest in using
the resulting data. We are very grateful to the referee, Dr. Jeffrey L.
Linsky, for his useful suggestions and comments that have led to
substantial improvements in the paper. I.~R. acknowledges support from the
Spanish Ministerio de Ciencia y Tecnolog\'{\i}a through a Ram\'on y Cajal
fellowship.  We acknowledge with gratitude the support for the ``Sun in
Time'' program from NASA grants NAG 5-382 (IUE), NAG 5-1662 (ROSAT), NAG
5-1703 (IUE), NAG 5-2160 (IUE), NAG 5-2494 (ROSAT), NAG 5-2707 (ASCA), NAG
5-3136 (EUVE), NAG 5-8985 (FUSE), NAG 5-10387 (FUSE), and NAG 5-12125
(FUSE). General stellar X-ray astronomy research at PSI has been supported
by the Swiss National Science Foundation under projects 2100-049343.96,
20-58827.99, and 20-66875.01.  This research has made use of NASA's
Astrophysics Data System.


\begin{deluxetable}{lrlrlccclcll}
\tablewidth{0pt}
\tablefontsize{\scriptsize}
\tablecaption{Relevant data for the studied Sun in Time targets
\label{tabtarg}}
\tablehead{\colhead{Name} &
\colhead{HD} &
\colhead{Sp.} &
\colhead{d} &
\colhead{$N_{\rm H}$} &
\colhead{$T_{\rm eff}$} &
\colhead{Mass} &
\colhead{Radius} &
\colhead{Sun R\tablenotemark{a}} &
\colhead{$P_{\rm rot}$} &
\colhead{Age} &
\colhead{Age}\\
\colhead{} &
\colhead{} &
\colhead{Typ.} &
\colhead{(pc)} &
\colhead{(cm$^{-2}$)} &
\colhead{(K)} &
\colhead{(M$_{\odot}$)} &
\colhead{(R$_{\odot}$)} &
\colhead{(R$_{\odot}$)} &
\colhead{(d)} &
\colhead{(Gyr)} &
\colhead{indicator}}
\startdata
EK Dra        & 129333 & G1.5~V&34.0 &$\sim1.5\times10^{18}$&5870&1.06&0.95&0.900&2.68    &0.1    & Local Association \& Li\\
$\pi^1$ UMa   &  72905 & G1.5~V&14.3 &$\sim1\times10^{18}$&5850&1.03&0.95&0.902&4.90    &0.3    & UMa group\\
$\chi^1$ Ori  &  39587 & G1~V  & 8.7 &$9\times10^{17}$&5890&1.01&0.96&0.902&5.24    &0.3    & UMa group\\
$\kappa^1$ Cet&  20630 & G5~V  & 9.2 &$8\times10^{17}$&5750&1.02&0.93&0.910&9.21    &0.65   & $P_{\rm rot}$-age \& L$_{\rm X}$\\
$\beta$ Com   & 114710 & G0~V  & 9.2 &$\sim1\times10^{18}$&6000&1.10&1.08&0.925&12      &1.6    & $P_{\rm rot}$-age rel.\\
Sun           &  --    & G2~V  & 1 AU& 0                &5777&1.00&1.00&1.00 &25.4    &4.6    & Isotopic dating\\
$\beta$ Hyi   &   2151 & G2~IV & 7.5 &$\sim2\times10^{18}$&5774&1.10&1.90&1.10 &$\sim$28&6.7    & Isochrones\\
\enddata
\tablenotetext{a}{Radius of the Sun at the same age from the models of Girardi
et al. (2000).}
\end{deluxetable}

\begin{deluxetable}{lrrrl}
\tablewidth{0pt}
\tablefontsize{\footnotesize}
\tablecaption{Summary of space missions used in this investigation
\label{tabobs}}
\tablehead{\colhead{Instrument}&
\colhead{Wavelength range (\AA)}&
\colhead{Calibration method}}
\startdata
ASCA                   & 1--40                          & Multi-$T_e$ plasma model \\
ROSAT                  & 6--124                         & Multi-$T_e$ plasma model \\
EUVE                   & 80--760\tablenotemark{\dagger} & Flux calibrated \\
FUSE                   & 920--1180                      & Flux calibrated \\
HST                    & 1150--1730                     & Flux calibrated \\
IUE                    & 1150--1950\tablenotemark{\dagger\dagger} & Flux calibrated \\
\enddata
\tablenotetext{\dagger}{$\lambda$$>$360 \AA\ not useful because of strong 
interstellar absorption.}
\tablenotetext{\dagger\dagger}{Used the SWP camera and for $\lambda$$<$1700
\AA\ only.}
\end{deluxetable}

\begin{deluxetable}{lcccccc}
\tablewidth{0pt}
\tablefontsize{\footnotesize}
\tablecaption{Observation dates and ids for all datasets used in this
investigation
\label{tabobsid}}
\tablehead{\colhead{Target}&
\multicolumn{2}{c}{ASCA}&
\multicolumn{2}{c}{ROSAT\tablenotemark{a}}&
\multicolumn{2}{c}{EUVE}\\
 & 
\colhead{Date}& \colhead{Obs ID}&
\colhead{Date}& \colhead{Obs ID}&
\colhead{Date}& \colhead{Obs ID}}
\startdata
EK Dra        & 1994-05-24 & 22012000 & 1993-10-19 & rp201474n00 & 1995-12-06 & ek\_dra\_\_9512061129   \\
$\pi^1$ UMa   & 1993-11-13 & 21018000 & 1993-10-05 & rp201472n00 & 1998-11-30 & 3\_uma\_\_9811301325    \\
              &            &          &            &             & 1998-12-05 & 3\_uma\_\_9812050029    \\
$\chi^1$ Ori  &    --      &    --    &    --      &     --      & 1993-01-26 & chi1\_ori\_\_9301261159 \\
$\kappa^1$ Cet& 1994-08-16 & 22013000 & 1993-07-27 & rp201473n00 & 1994-10-13 & kappa\_cet\_\_9410131500\\
              &            &          &            &             & 1995-10-06 & kappa\_cet\_\_9510061036\\
$\beta$ Com   &    --      &    --    & 1993-06-17 & rp201471n00 &    --      &           --            \\
$\beta$ Hyi   &    --      &    --    & 1991-04-21 & rp200071n00 &    --      &           --            \\
\hline
              &            &          &            &             &            &                         \\
\hline
\hline
Target        &\multicolumn{2}{c}{FUSE}&\multicolumn{2}{c}{HST}& \multicolumn{2}{c}{IUE}\\
              &    Date    & Obs ID   &   Date     &    Obs ID &     Date   & Obs ID    \\
\hline
EK Dra        & 2002-05-14 & C1020501 &    --      &     --    & 1992-05-31 & SWP 44817 \\
$\pi^1$ UMa   & 2001-05-12 & B0780101 &    --      &     --    & 1980-03-28 & SWP 08582 \\
              &            &          &            &           & 1990-10-12 & SWP 39813 \\
$\chi^1$ Ori  &     --     &    --    & 2000-03-10 & o5bn02010 & 1990-02-01 & SWP 38108 \\
              &            &          & 2000-03-10 & o5bn02020 & 1984-04-03 & SWP 22408 \\
$\kappa^1$ Cet& 2000-09-10 & A0830301 & 2000-09-19 & o5bn03050 & 1994-09-14 & SWP 52115 \\
              &            &          & 2000-09-19 & o5bn03060 & 1994-08-16 & SWP 51829 \\
              &            &          &            &           & 1994-08-16 & SWP 51831 \\
$\beta$ Com   & 2001-01-26 & A0830401 &    --      &     --    & 1982-02-11 & SWP 16313 \\
              &            &          &            &           & 1979-08-14 & SWP 06179 \\
$\beta$ Hyi   & 2000-07-01 & A0830101 &    --      &     --    & 1979-12-18 & SWP 07430 \\
              &            &          &            &           & 1979-12-03 & SWP 07307 \\
              &            &          &            &           & 1994-05-14 & SWP 50765 \\
              &            &          &            &           & 1992-03-13 & SWP 44168 \\
\enddata
\tablenotetext{a}{EK Dra, $\chi^1$ Ori, $\kappa^1$ Cet, and $\beta$ Hyi have
ROSAT observations using the Boron filter but were not used here.}
\end{deluxetable}

\begin{deluxetable}{lcccccc}
\tablewidth{0pt}
\tablefontsize{\footnotesize}
\tablecaption{Integrated fluxes (in units of erg s$^{-1}$ cm$^{-2}$) normalized
to a distance of 1 AU and to the radius of a one solar mass star. See text for 
details
\label{tabfltot}}
\tablehead{
\colhead{}              &
\colhead{0.10 Gyr}      &
\colhead{0.30 Gyr}      &
\colhead{0.65 Gyr}      &
\colhead{1.6 Gyr}       &
\colhead{4.56 Gyr}      &
\colhead{6.7 Gyr}\\
\colhead{$\lambda$ interval (\AA)}&
\colhead{(EK Dra)}                &
\colhead{($\pi^1$ UMa + $\chi^1$ Ori)}  &
\colhead{($\kappa^1$ Cet)}        &
\colhead{($\beta$ Com)}           &
\colhead{(Sun)}                   &
\colhead{($\beta$ Hyi)}}
\startdata
$\left[1-20\right]$                        &  \phantom{0}180.2\phantom{0}\phantom{0}            &  \phantom{0}\phantom{0}21.5\phantom{0}\phantom{0}    &   \phantom{0}\phantom{0}\phantom{0}7.76      &  \phantom{0}\phantom{0}0.976\phantom{0}             &   0.15     &  \phantom{0}\phantom{0}0.048\phantom{0}                        \\ 
$\left[20-100\right]$                      &  \phantom{0}\phantom{0}82.4\phantom{0}\phantom{0}  &  \phantom{0}\phantom{0}14.8\phantom{0}\phantom{0}    &   \phantom{0}\phantom{0}10.7\phantom{0}      &  \phantom{0}\phantom{0}2.80\phantom{0}\phantom{0}   &   0.70     &  \phantom{0}\phantom{0}0.321\phantom{0}                        \\
$\left[100-360\right]$                     &  \phantom{0}187.2\phantom{0}\phantom{0}            &  \phantom{0}\phantom{0}69.4\phantom{0}\phantom{0}    &   \phantom{0}\phantom{0}22.7\phantom{0}      &  \phantom{0}(7.7)\phantom{0}\phantom{0}             &   2.05     &  \phantom{0}(1.37)\phantom{0}                                  \\
$\left[360-920\right]$                     &  ((45.6))                                          &  ((15.2))                                            &   ((7.0))                                    &  ((2.85))                                           &   1.00     &                                                    ((0.68))    \\
$\left[920-1180\right]$                    &  \phantom{0}(18.1)\phantom{0}                      &  \phantom{0}\phantom{0}\phantom{0}8.38\phantom{0}    &   \phantom{0}\phantom{0}2.90\phantom{0}      &  \phantom{0}(1.70)\phantom{0}                       &   0.74     &  \phantom{0}(0.50)\phantom{0}                                  \\
$\left[1-360\right]+\left[920-1180\right]$ &  \phantom{0}467.9\phantom{0}\phantom{0}            &  \phantom{0}114.1\phantom{0}\phantom{0}              &   \phantom{0}44.1\phantom{0}\phantom{0}      &  \phantom{0}13.2\phantom{0}\phantom{0}\phantom{0}   &   3.64     &  \phantom{0}\phantom{0}2.2\phantom{0}\phantom{0}\phantom{0}    \\
$\left[1-1180\right]$                      &  \phantom{0}513.5\phantom{0}\phantom{0}            &  \phantom{0}129.3\phantom{0}\phantom{0}              &   \phantom{0}51.1\phantom{0}\phantom{0}      &  \phantom{0}16.0 \phantom{0}\phantom{0}             &   4.64     &  \phantom{0}\phantom{0}2.9 \phantom{0}\phantom{0}              \\
\enddata
\end{deluxetable}

\begin{deluxetable}{lrc}
\tablewidth{0pt}
\tablecaption{Parameters of the power-law fits to the measured integrated
fluxes\tablenotemark{a}
\label{tabslop}}
\tablehead{
\colhead{$\lambda$ interval (\AA)}&
\colhead{$\alpha$}&
\colhead{$\beta$}                   }
\startdata
$\left[1-20\right]$                        & 2.4\rlap{0} & \phantom{(}$-$1.92  \\
$\left[20-100\right]$                      & 4.4\rlap{5} & \phantom{(}$-$1.27  \\
$\left[100-360\right]$                     & 13.5        & \phantom{(}$-$1.20  \\
$\left[360-920\right]$                     & 4.5\rlap{6} &           ($-$1.0)  \\
$\left[920-1180\right]$                    & 2.5\rlap{3} & \phantom{(}$-$0.85  \\
$\left[1-360\right]+\left[920-1180\right]$ & 24.8        & \phantom{(}$-$1.27  \\
$\left[1-1180\right]$                      & 29.7        & \phantom{(}$-$1.23  \\
\enddata
\tablenotetext{a}{Relationship of the form: $\mbox{Flux}=\alpha \: \mbox{[age
(Gyr)]}^{\beta}$.} 
\end{deluxetable}

\begin{deluxetable}{rlcccccc}
\tablewidth{0pt}
\tablefontsize{\footnotesize}
\tablecaption{Integrated fluxes (in units of erg s$^{-1}$ cm$^{-2}$) of strong
emission features normalized to a distance of 1 AU and the radius of a one
solar mass star. See text for details
\label{tabflin}}
\tablehead{
\colhead{}             &
\colhead{}             &
\colhead{0.10 Gyr}     &
\colhead{0.30 Gyr}     &
\colhead{0.65 Gyr}     &
\colhead{1.6 Gyr}      &
\colhead{4.56 Gyr}     &
\colhead{6.7 Gyr}\\
\colhead{$\lambda$ (\AA)} &
\colhead{main elem.}      &
\colhead{(EK Dra)}        &
\colhead{($\pi^1$ UMa + $\chi^1$ Ori)}   &
\colhead{($\kappa^1$ Cet)}&
\colhead{($\beta$ Com)}   &
\colhead{(Sun)}           &
\colhead{($\beta$ Hyi)}}
\startdata
284      &Fe {\sc xv}  &           22.0 & 5.0\phantom{0} & 2.4\phantom{0}         &  --   & 0.025            &  --   \\
304      &He {\sc ii}  &           44.3 & 8.3\phantom{0} & 2.3\phantom{0}         &  --   & 0.260            &  --   \\
335      &Fe {\sc xvi} &           36.6 & 9.7\phantom{0} & 2.6\phantom{0}         &  --   &  --              &  --   \\
361      &Fe {\sc xvi} &           15.7 & 6.6\phantom{0} & 1.6\phantom{0}         &  --   & 0.016            &  --   \\
584      &He {\sc i}   &            --  &  --            &  --                    &  --   & 0.032            &  --   \\
610\&625 &Mg {\sc x}   &            --  &  --            &  --                    &  --   & 0.028            &  --   \\
630      &O {\sc v}    &            --  &  --            &  --                    &  --   & 0.037            &  --   \\
789      &O {\sc iv}   &            --  &  --            &  --                    &  --   & 0.017            &  --   \\
834      &O {\sc ii}   &            --  &  --            &  --                    &  --   & 0.015            &  --   \\
977      &C {\sc iii}  & \phantom{1}5.0 & 1.22           & 0.59                   & 0.30  & 0.150            & 0.124 \\
1026     &H {\sc i}    &            --  & 3.1\phantom{5} & 0.80                   &  --   & 0.098            &  --   \\
1032     &O {\sc vi}   & \phantom{1}3.1 & 0.75           & 0.43                   & 0.16  & 0.050            & 0.048 \\
1038     &O {\sc vi}   & \phantom{1}1.5 & 0.38           & 0.21                   & 0.074 & 0.025            & 0.022 \\
1176     &C {\sc iii}  & \phantom{1}3.4 & 0.73           & 0.37                   & 0.15  & 0.053            & 0.046 \\
1206     &Si {\sc iii} &            --  & 1.12           & 0.75                   &  --   & 0.095            &  --   \\
1216     &H {\sc i}    &            --  & \llap{4}2.2\phantom{0}    &\llap{2}9.3\phantom{0}  &  --   & 6.19\phantom{0}  &  --   \\
1304     &O {\sc i}    & \phantom{1}4.3 & 1.18           & 0.60                   & 0.45  & 0.143            & 0.163 \\
1335     &C {\sc ii}   & \phantom{1}4.7 & 1.52           & 0.95                   & 0.36  & 0.188            & 0.155 \\
1400     &Si {\sc iv}  & \phantom{1}4.3 & 1.59           & 0.77                   & 0.28  & 0.083            & 0.097 \\
1550     &C {\sc iv}   & \phantom{1}9.1 & 2.21           & 1.02                   & 0.40  & 0.146            & 0.082 \\
1640     &He {\sc ii}  & \phantom{1}6.0 & 0.99           & 0.56                   &  --   & 0.040            &  --   \\
1657     &C {\sc i}    & \phantom{1}4.1 & 0.97           & 0.78                   & 0.47  & 0.202            & 0.210 \\
\enddata
\end{deluxetable}

\begin{deluxetable}{rlcc}
\tablewidth{0pt}
\tablefontsize{\footnotesize}
\tablecaption{Ion formation temperatures and slopes of the power-law fits to
the measured line fluxes in Table \ref{tabflin}
\label{tabslin}}
\tablehead{
\colhead{$\lambda$ (\AA)}&
\colhead{main elem.}     &
\colhead{$\log T$}       &
\colhead{Slope}}
\startdata
284      &Fe {\sc xv}  & 6.30 & $-$1.79\\
304      &He {\sc ii}  & 4.75 & $-$1.34\\
335      &Fe {\sc xvi} & 6.35 & -- \\
361      &Fe {\sc xvi} & 6.35 & $-$1.86\\
584      &He {\sc i}   & 4.25 & -- \\
610\&625 &Mg {\sc x}   & 6.08 & -- \\
630      &O {\sc v}    & 5.26 & -- \\
789      &O {\sc iv}   & 5.05 & -- \\
834      &O {\sc ii}   & 4.80 & -- \\
977      &C {\sc iii}  & 4.68 & $-$0.85\\
1026     &H {\sc i}    & 3.84 & $-$1.24\\
1032     &O {\sc vi}   & 5.42 & $-$1.00\\
1038     &O {\sc vi}   & 5.42 & $-$1.02\\
1176     &C {\sc iii}  & 4.68 & $-$1.02\\
1206     &Si {\sc iii} & 4.40 & $-$0.94\\
1216     &H {\sc i}    & 3.84 & $-$0.72\\
1304     &O {\sc i}    & 3.85 & $-$0.78\\
1335     &C {\sc ii}   & 4.25 & $-$0.78\\
1400     &Si {\sc iv}  & 4.75 & $-$0.97\\
1550     &C {\sc iv}   & 5.00 & $-$1.08\\
1640     &He {\sc ii}  & 4.75 & $-$1.28\\
1657     &C {\sc i}    & 3.85 & $-$0.68\\
\enddata
\end{deluxetable}

\end{document}